\documentclass[aps,prb,twocolumn,showpacs]{revtex4}
\usepackage{graphics,amsmath}
\usepackage{tabularx,graphicx}
\usepackage[latin1]{inputenc}
\usepackage{epsfig}

\newcommand{\alf}{\alpha_b}
\newcommand{\e}{\varepsilon}

\newcommand{\D}{\Delta}

\newcommand{\Dr}{\Delta_{r}}

\newcommand{\Dz}{\Delta_z}

\newcommand{\adk}{a^\dag_k}
\newcommand{\om}{\omega}

\newcommand{\wco}{\omega_{co}}
\newcommand{\wo}{\omega_{0}}

\newcommand{\Dox}{\Delta^x_0}

\newcommand{\Doz}{\Delta^z_0}

\begin{document}
\title{Surface dissipation in nanoelectromechanical systems: Unified description with the Standard Tunneling Model, and effects of metallic electrodes.}
\author{C. Seo\'anez}
\affiliation{Instituto de  Ciencia de Materiales de Madrid, CSIC,
 Cantoblanco E28049 Madrid, Spain}

\author{F. Guinea}
\affiliation{Instituto de  Ciencia de Materiales de Madrid, CSIC,
 Cantoblanco E28049 Madrid, Spain}

\author{A.~H. Castro Neto}
\affiliation{Department of Physics, Boston University, 590
Commonwealth Avenue, Boston, MA 02215, USA}


\begin{abstract}
Modifying and extending recent ideas $\,\,$\cite{SGN07}, a
theoretical framework to describe dissipation processes in the
surfaces of vibrating micro and nanoelectromechanical devices
(MEMS-NEMS), thought to be the main source of friction at low
temperatures, is presented. Quality factors as well as frequency
shifts of flexural and torsional modes in doubly-clamped beams and
cantilevers are given, showing the scaling with dimensions,
temperature and other relevant parameters of these systems. Full
agreement with experimental observations is not obtained, leading to
a discussion of limitations and possible modifications of the scheme
to reach quantitative fitting to experiments. For NEMS covered with
metallic electrodes the friction due to electrostatic interaction
between the flowing electrons and static charges in the device and
substrate is also studied.
\end{abstract}


\maketitle
\section{Introduction}
The successful race for miniaturization of semiconductor
technologies $\,\,$\cite{S81} manifests itself spectacularly in the
form of nanoelectromechanical systems
(NEMS)$\,\,$\cite{C00,C02,B04,ER05}, machines in the micron and
submicron scale whose mechanical motion, integrated into electrical
circuits, has a wealth of technological applications, including
control of currents at the single-electron level $\,\,$\cite{SB04},
single-spin detection $\,\,$\cite{RBMC04}, sub-attonewton force
detection $\,\,$\cite{MR01}, mass sensing of individual molecules
$\,\,$\cite{EHR04}, high-precision thermometry
$\,\,$\cite{hopcroft:013505} or \textit{in-vitro} single-molecule
biomolecular recognition $\,\,$\cite{DKEM06}.

\begin{figure}
\begin{center}
\includegraphics[width=7cm]{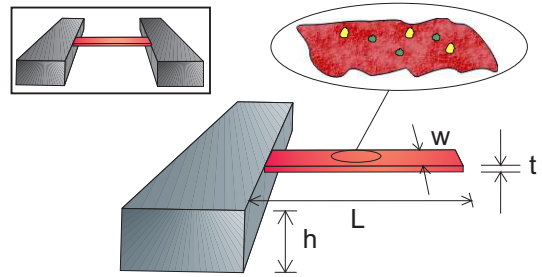}\\
\caption[fig]{\label{sketch_resonator} (Color online) Sketch of the
systems considered in the text. The inset shows a doubly clamped
beam, while the main figure shows a cantilever characterized by its
dimensions, width ($w$), thickness ($t$), and length, ($L$), where
$w \sim t \ll L$. The height above the substrate is $h$. A schematic
view of the surface is given, highlighting imperfections like
roughness and adsorbates, which dominate dissipation at low
temperatures.}
\end{center}
\end{figure}

These mechanical elements (like cantilevers or beams, see
fig.[\ref{sketch_resonator}]) are also a focus of attention and
intensive research as experiments are approaching the quantum regime
$\,\,$\cite{LBCS04,ZGBM05,Netal06}, where manifestations of
quantized mechanical motion of a macroscopic degree of freedom like
their center of mass should become apparent. Several schemes to
prepare the mechanical oscillator in a non-classical state and
observe clear signatures of its quantum behavior have been recently
suggested
$\,\,$\cite{santamore:144301,martin:120401,jacobs:147201,WLSN06,Tetal07,katz:040404}.

\begin{figure}
\begin{center}
\includegraphics[width=7cm]{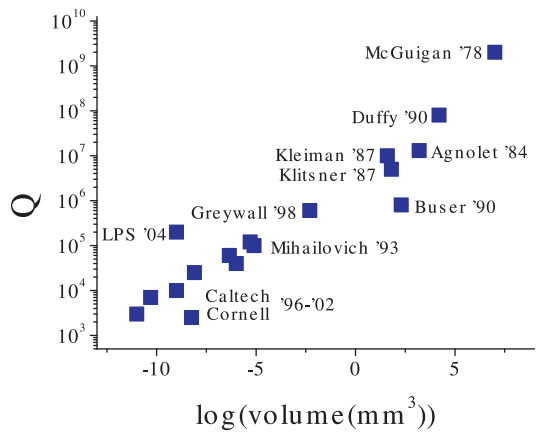}\\
\caption[fig]{\label{Evolution_size} (Color online) From ref.
$\,\,$\cite{ER05}. Evolution of reported quality factors in
monocrystalline mechanical resonators with size, showing a decrease
with linear dimension, i.e., with increasing surface-to-volume
ratio, indicating a dominant role of surface-related losses.}
\end{center}
\end{figure}

A key figure of merit of the mechanical oscillation is its quality
factor $Q=\om/\Delta\om$, where $\Delta\om$ is the measured
linewidth of the corresponding vibrational eigenmode of frequency
$\om$. To reach the quantum regime, as well as for most practical
applications, where the measured shifts of the resonant frequency
$\om$ constitute the detectors principle, a very high $Q$ is
compulsory. Imperfections and the environment surrounding the
oscillator result in both a finite linewidth $\Delta\om$ and a
frequency shift $\delta\om$ with respect to the ideal case.
Therefore several works have been devoted to the analysis of the
different sources of dissipation present in MEMS and NEMS
$\,\,$\cite{CR99,Yetal00,Eetal00,CR02,YOE02,Metal02,AM03,Hetal03,ZGSBM05,FZMR06,Mohetal07},
trying to determine the dominant damping mechanisms and ways to
minimize them. Among the different mechanisms affecting
semiconductor-based NEMS the most important and difficult to avoid
are i) clamping losses $\,\,$\cite{JI68,PJ04}, through the transfer
of energy from the resonator mode to acoustic modes at the contacts
and beyond to the substrate, ii) thermoelastic damping
$\,\,$\cite{Z38,LR00,de:144305} and iii) friction processes taking
place at the surfaces $\,\,$\cite{YOE00,LVSLHP05,C07}. At low
temperatures and for decreasing sizes the prevailing mechanism is
the last one $\,\,$\cite{ER05}, as indicated by the linear decrease
of the quality factor of flexural modes with decreasing size (see
fig.[\ref{Evolution_size}]), or the sharp increase of $Q$ when the
resonator is annealed $\,\,$\cite{YOE00,WOE04}. Excitation of
adsorbed molecules, movement of lattice defects or configurational
rearrangements absorb irreversibly energy from the excited eigenmode
and redistribute it among the rest of degrees of freedom of the
system.

A theoretical quantitatively accurate description of surface
dissipation proves therefore challenging, as many different
dynamical processes and actors come into play, some of whom are not
yet well characterized, so simplifications need to be made to
provide a unified framework for all of them. In $\,\,$\cite{SGN07}
such a scheme was given, based on the following considerations: i)
experimental observations indicate that surfaces of otherwise
monocrystalline resonators acquire a certain degree of roughness,
impurities and disorder, resembling an amorphous structure
$\,\,$\cite{LTWP99}. ii)In amorphous solids the damping of acoustic
waves at low temperatures is successfully explained by the Standard
Tunneling Model $\,\,$\cite{AHV72,P72,P87,E98}, which couples the
acoustic phonons to a set of Two-Level Systems (TLSs) representing
the low-energy spectrum of all the degrees of freedom (DoF) able to
exchange energy with the strain field associated to the vibration.
These DoF correspond to impurities or clusters of atoms within the
structure which have in their configurational space two energy
minima separated by an energy barrier (similarly to the dextro/levo
configurations of the ammonia molecule), modeled as a DoF tunneling
between two potential wells. At low temperatures only the two lowest
eigenstates have to be considered, characterized by the bias $\Doz$
between the wells and the tunneling rate $\Dox$ through the barrier.
The Standard Tunneling Model specifies the properties of the set of
TLSs in terms of a probability distribution $P(\Dox,\Doz)$ which can
be inferred from general considerations $\,\,$\cite{AHV72,P72} and
is supported by experiments $\,\,$\cite{E98}. It has to be noted,
however, that below a certain temperature the model breaks down due
to the increasing role played by interactions among the TLSs, not
included $\,\,$\cite{YL88,ERK04}.

In $\,\,$\cite{SGN07} a description of the attenuation of vibrations
in nanoresonators due to their amorphous-like surfaces in terms of
an adequate adaptation of the Standard Tunneling Model was intended
to be given. An estimate for $Q^{-1}(T)$ was provided, reproducing
correctly the weak $\sim T^{1/3}$ temperature dependence as well as
the order of magnitude observed in recent experiments
$\,\,$\cite{ZGSBM05,Mohetal07}. However, the way the Standard
Tunneling Model was adapted was not fully correct, and a revision of
the results for $Q^{-1}(T)$ obtained therein was mandatory.

In this work we will i) Discuss in detail some important issues
hardly mentioned in $\,\,$\cite{SGN07}, ii) Modify some points to
obtain a theory fully consistent with the Standard Tunneling Model
and which includes all possible dissipative processes due to the
presence of TLSs, iii) Extend the results and give expressions for
the quality factor of cantilevers as well as damping of torsional
modes, and frequency shifts associated, iv) Compare with available
experimental data, discussing the validity of the model, range of
applicability, aspects to be modified/included to reach an accurate
quantitative fit to experiments, v) Study the dissipative effects
associated to the presence of metallic electrodes frequently
deposited on top of the resonators.

Section II starts with a brief summary of the model described in
$\,\,$\cite{SGN07}, discussing the approximations involved, while
the different dissipative mechanisms due to the presence of TLSs are
presented in III. Section IV analyzes and compares the two main
mechanisms, namely relaxational processes associated to asymmetric
(biased) TLSs, and non-resonant damping of symmetric TLSs. The
extension of the results to cantilevers and torsional modes is given
in Section V. A brief account of frequency shifts' expressions is
found in Section VI. In Section VII comparison with experiments and
discussion of the applicability, validity and further extensions of
the model is made. Section VIII discusses dissipation effects due to
the existence of metallic leads coupled to the resonator, which can
be coupled to the electrostatic potential induced by trapped charges
in the device. Under some circumstances, these effects cannot be
neglected. Finally, the main conclusions of our work are given in
Section IX. Details of some calculations are provided in the
Appendices.

We will not analyze here dissipation due to clamping and
thermoelastic losses, which may dominate dissipation in the case of
very short beams and strong driving, respectively. We will also not
consider direct momentum exchange processes between the carriers in
the metallic circuit and the vibrating system$\,\,$\cite{SLB02}.

\section{Surface friction modeled by TLS$\rm{s}$}
\subsection{Hamiltonian}
We will consider a rod of length $L$, width $w$ and thickness $t$,
see fig.[\ref{sketch_resonator}], and use units such that
$\hbar=1=k_B$. As described in $\,\,$\cite{SGN07}, the vibrating
resonator with imperfect surfaces is represented by its vibrational
eigenmodes coupled to a collection of non-interacting TLSs, assuming
that the main effect of the strain caused by the phonons is to
modify the bias $\Doz$ of the TLSs $\,\,$\cite{A86}:
\begin{eqnarray}\label{H1}
    H&=&\sum_{k,j}\om_{k,j}a_{k,j}^\dag
a_{k,j}+\\
\nonumber&+&\sum_{\Dox,\Doz}\Bigl\{\Dox\sigma_x+\Bigl[\Doz+\sum_{k,j}
\lambda_{k,j} \left( b_{k,j}^\dag +
b_{k,j}\right)\Bigr]\sigma_z\Bigr\}
\end{eqnarray}
The index $j$ represents the three kinds of modes present in a thin
beam geometry $\,\,$\cite{LL59}: flexural (bending), torsional and
compression modes. These modes will be present for wavelengths
$\lambda>t$, while for shorter ones the system is effectively 3D,
with the corresponding 3D modes. The main effect of these high
energy 3D modes is to renormalize the tunneling amplitude, so we
will take that renormalized value as our starting point $\Dox$ and
forget in the following about the 3D modes. The sum over TLSs is
characterized by the probability distribution
$P(\Dox,\Doz)=P_0/\Dox$ . $\,\,$\cite{AHV72,P72} This result
follows, as explained in $\,\,$\cite{P87,P72}, just due to i) the
exponential sensitivity of $\Dox$ to the properties of the energy
barrier of the two-well potential giving rise at low T to the TLS
description of the system (resulting in a $1/\Dox$ dependence of
$P(\Dox,\Doz)$), and ii) the characteristic energy scale of the
distribution of asymmetries $\Doz$, much bigger than 1K, which is
the temperature at which experiments are performed, and which thus
fixes the scale of the bias of a TLS if it is to contribute
significantly to dissipation, $\Doz\leq1$K (therefore the relevant
TLSs have values of $\Doz$ lying in a very narrow energy range
around $\Doz=0$ as compared to the variance of their probability
distribution, allowing us to consider that $P(\Dox,\Doz)$ to a first
approximation does not depend on $\Doz$). Unphysical divergencies do
not appear, as $\Dox>\Delta_{min}$, with $\Delta_{min}$ fixed by the
typical timescale of the experiment, given by the time needed to
obtain a spectrum around the resonance frequency of the excited
vibrational eigenmode of the resonator, and
$\e=\sqrt{(\Dox)^2+(\Doz)^{2}}<\e_{max}$, estimated to be of the
order of 5 K $\,\,$\cite{E98}. For typical amorphous insulators
$P_0\sim10^{44}$J$^{-1}$m$^{-3}$.

To see to which low energy modes the TLSs are more coupled, inducing
a more effective dissipation at low temperatures, the spectral
function $ J ( \omega ,j) \equiv \sum_k \left| \lambda_{k,j}
\right|^2 \delta ( \omega - \omega_{k,j} ) $ characterizing the
evolution of the strength of the coupling for each type of mode can
be computed. Due to their nonlinear dispersion relation,
$\om=\sqrt{EI/(\rho tw)} k^2$, with $I = t^3 w/12$, $E$ the Young
modulus and $\rho$ the mass density, flexural modes show a subohmic
behavior $J_{{\rm flex}}(\omega) = \alpha_b\sqrt{\wco}
\sqrt{\omega}$, with
\begin{eqnarray}\label{Jsubohmic}
 \alpha_b \sqrt{\wco} =
 0.3\frac{\gamma^2}{t^{3/2}w}\frac{(1+\nu)(1-2\nu)}{E(3-5\nu)}
   \Bigl(\frac{\rho}{E}\Bigr)^{1/4} \, ,
\end{eqnarray}
where $\gamma\sim5$ eV is a coupling constant appearing in
$\lambda_{k,j}$, $\nu$ is Poisson's ratio and $\wco\simeq
\sqrt{EI/(\rho tw)}(2\pi/t)^2$ is the high energy cut-off of the
bending modes. A detailed derivation of $J_{{\rm flex}}(\omega)$ is
given in Appendix \ref{DerivJ}. Even though the length $L$ of our
system is finite, and thus the vibrational spectrum discrete, a
continuum approximation like this one will hold if
$kT\gg\hbar\om_{fund}$, $\om_{fund}$ being the frequency of the
fundamental mode.

The bending modes prevail over the other, ohmic-like, modes as a
dissipative channel at low energies, thanks to their weaker $J_{{\rm
flex}}(\omega)\sim\om^{1/2}$ dependence. One may ask at what
frequency do the torsional and compression modes begin to play a
significant role, and a rough way to estimate it is to see at what
frequency do the corresponding spectral functions have the same
value, $J_{{\rm flex}}(\om*)=J_{{\rm comp,tors}}(\om*)$. Using the
expressions in \cite{SGN07}, namely $J_{{\rm
comp,tors}}(\om)=\alpha_{c,t}\om$, with $\alpha_c= (\gamma
\Delta_{0}^{x}/\Delta_{0})^2 (2 \pi^2 \rho t w )^{-1}
(E/\rho)^{-3/2}$ and $\alpha_t = C (\gamma
\Delta_{0}^{x}/\Delta_{0})^2(8\pi^2 \mu t w  \rho I)^{-1} (\rho
I/C)^{3/2}$, the results are
$\om*\sim30(1+\nu)^2(1-2\nu)^2(E/\rho)^{1/2}/[t(3-5\nu)^2]$ for the
case of compression modes and
$\om*\sim300(1-2\nu)^2(E/\rho)^{1/2}/[t(3-5\nu)^2(1+\nu)]$ for the
torsional. Comparing these frequencies to the one of the onset of 3D
behavior, $\wco$, they are similar, justifying a simplified model
where only flexural modes are considered.

\subsection{TLSs dynamics}
\begin{figure}
\begin{center}
\includegraphics[width=7cm]{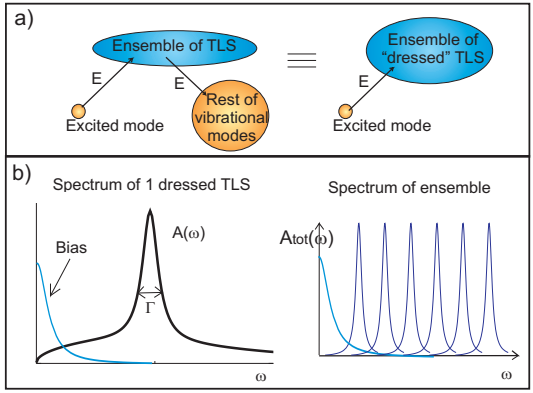}\\
\caption[fig]{\label{spectrum} (Color online)a) Schematic
representation of the irreversible flow of energy from the
externally excited mode to the TLS ensemble, and from the ensemble
to the rest of vibrational modes. This process can be viewed as a
flow of energy from the excited mode to an ensemble of "dressed"
TLSs, with their dynamics modified by the presence of the
vibrational modes. b) Left: Spectral function $A(\om)$ of a single
dressed TLS, weakly damped ($\Gamma<\Dox$). A peak around $\om=0$
arises if the system is biased, corresponding to the relaxational
dissipation mechanism. Right: Total spectral function $A_{tot}(\om)$
of the ensemble of dressed TLSs.}
\end{center}
\end{figure}

The interaction between the bending modes and the TLSs affects both
of them. When a single mode is externally excited, as is done in
experiments, the coupling to the TLSs will cause an irreversible
energy flow, from this mode to the rest of the modes through the
TLSs, as depicted in fig.(\ref{spectrum}a). The dynamics of the TLSs
in presence of the vibrational bath determines the efficiency of the
energy flow and thus the quality factor of the excited mode. Taking
a given TLS plus the phonons, its dynamics is characterized by the
Fourier transform of the correlator
$\langle\sigma_z(t)\sigma_z(0)\rangle$, the spectral function
$A(\om)$, which at $T=0$ reads
\begin{equation}
A ( \omega ) \equiv \sum_n \left| \left\langle 0 \left| \sigma_z
\right| n
  \right\rangle \right|^2 \delta ( \omega - \omega_n + \omega_0 )
\label{spectral_TLS}
\end{equation}
where $| n \rangle$ is an excited state of the total system TLS plus
vibrations. In $\,\,$\cite{SGN07} an analysis of $A(\om)$ was made
for the case of a \emph{symmetric} TLS ($\Doz=0$), concluding that
i)If $\Dox\ll\alf^2\wco$ the tunneling amplitude is basically
suppressed and the TLS does not participate in dissipative
processes. For reasonable system dimensions the coupling constant is
very small, $\alf\ll1$, so this effect can be ignored, ii)Around the
resonance at $\om=\Dox$ a broadening appears, $\Gamma(\Dox)$ which,
for $\Gamma(\Dox)\ll\Dox$ is given by the Fermi Golden Rule result
$\Gamma(\Dox)= 16 \alpha_b\sqrt{\wco} \, \sqrt{\Dox}$ (for $T=0$ K,
at $T>0$ eq.(\ref{gammaresonant}) applies), iii)The coupling to
phonons of all energies provides the "dressed" TLS with tails far
from resonance, $A ( \omega ) \propto \alpha_b \sqrt{\wco
\omega}/(\Dox)^2$ for $\om\ll\Dox$ and $A ( \omega ) \propto
\alpha_b \sqrt{\wco} (\Dox)^2 \omega^{-7/2}$, for  $\omega \gg
\Dox$, see left side of fig.(\ref{spectrum}b), iv) The main effect
of the asymmetry $\Doz$ is to suppress the TLSs dynamics, so the
TLSs playing an active role in dissipation satisfy $\Dox>|\Doz|$.

Finally, for those low-energy overdamped TLSs such that
$\Gamma(\Dox)\gg\Dox$ an incoherent decay $\exp(-t/\tau)$ with time,
with $\tau(\Dox)=\Gamma(\Dox)^{-1}$ was assumed.

Two points that ref.$\,\,$\cite{SGN07} missed, and now will be
considered, are the following: First, the effect of the asymmetry is
not simply the one stated. An \emph{underdamped biased} TLS develops
an additional peak around $\om=0$, as shown in
fig.(\ref{spectrum}b), which corresponds to the relaxation mechanism
that dominates dissipation of acoustic waves in amorphous solids
$\,\,$\cite{E98}, see eq.(\ref{QEsq}). In $\,\,$\cite{SGN07} the
relaxation mechanism associated to eq.(\ref{QEsq}) was
misinterpreted to correspond to friction due to overdamped TLSs,
while in this work it will be, consistently with the Standard
Tunneling Model, linked to the presence of biased underdamped TLSs.
It will be described in more detail in the next section, and taken
into account in the computation of the total dissipation.

Second point missed by ref.$\,\,$\cite{SGN07}: one can estimate,
using the probability distribution $P(\Dox,\Doz)$, the total number
of overdamped TLSs in the volume fraction of the resonator
presenting amorphous features, $V_{amorph}$. With $V_{amorph}\sim
V_{tot}/10$, and using
$\Gamma(E,T)=\Gamma(E,T=0)\coth[E/2T]\approx2T\Gamma(E,T=0)/E$, the
number of overdamped TLSs, $\Dox \leq \Gamma(\Dox,T)\rightarrow \Dox
\leq [30 \alf\sqrt{ \wco} T]^{2/3}$, is $N\approx P_0twL[30
\alf\sqrt{ \wco} T]^{2/3}$, which for typical resonator sizes
$L\sim1\mu$m, $t,w\sim0.1\mu$m is less than one for $T<1$K.
Therefore, unless the resonator is bigger and/or $P_0$ too, this
contribution to dissipation can be safely neglected, as it will be
done in the following.

\section{Dissipative mechanisms}
From the features of the spectral function
$A_{tot}(\om)=\sum_{\Dox,\Doz}A(\Dox,\Doz,\om)$ of the ensemble of
dressed TLSs, one can classify into three kinds the dissipative
mechanisms affecting an externally excited mode $\wo$:
\subsection{Resonant dissipation}
Those TLSs with their unperturbed excitation energies close to $\wo$
will resonate with the mode, exchanging energy quanta, with a rate
proportional to the mode's phonon population $n_{\wo}$ to first
order. For usual excitation amplitudes $\sim1${\AA} the vibrational
mode is so populated (as compared with the thermal population) that
the resonant TLSs become saturated, and their contribution to the
transverse (flexural) wave attenuation becomes negligible,
proportional to $n_{\wo}^{-1/2}\,\,\,\,$ $\,\,$\cite{Aetal74}.
\subsection{Dissipation of symmetric non-resonant TLSs}
We show first that a correct description of the dissipation due to
weakly damped TLSs is given by the approximate Hamiltonian used in
$\,\,$\cite{SGN07}, starting from eq.(\ref{H1}), if one considers
apart in some way the presence of the relaxational peak at $\om=0$
due to a finite bias $\Doz\neq0$.

We focus the attention on a given TLS plus the vibrations
(spin-boson model),
$\textsl{H}=\Dox\sigma_x+\Doz\sigma_z+H_{int}+H_{vibr}$,
\begin{equation}\label{interaction}
    H_{int}=\sigma_z\sum_k \lambda\frac{k^2}{\sqrt{\omega_k}}(\adk
+a_k)
\end{equation}
with $\lambda$ defined in eq.(\ref{deflambda}). In
$\,\,$\cite{SGN07} a change of basis to the unperturbed eigenstates
of the TLS was performed, obtaining
$\textsl{H}=\e\sigma_z+[(\Dox/\e)\sigma_x+(\Doz/\e)\sigma_z]\sum_k
\lambda\frac{k^2}{\sqrt{\omega_k}}(\adk +a_k)+H_{vibr}$. Then the
term $(\Doz/\e)\sigma_z$ was ignored (remember
$\e=\sqrt{(\Dox)^2+(\Doz)^{2}}$), arguing that the key role in
dissipation is played by fairly symmetrical TLSs, $|\Doz|\ll\e$.
This allowed to simplify the spin-boson Hamiltonian to the one of a
symmetric TLS, much easier to analyze, with tunneling amplitude $\e$
instead of $\Dox$ and coupling $\lambda(\Dox/\e)$ instead of
$\lambda$. The consistency was kept by restricting the sums over the
TLS ensemble to those such that $|\Doz|\leq\Dox$.

One can check this consistency going back to the original basis,
where the approximation of the Hamiltonian reads
$\textsl{H}\approx\Dox\sigma_x+\Doz\sigma_z+[(\Dox/\e)^2\sigma_z-(\Doz\Dox/\e^2)\sigma_x]\sum_k
\lambda\frac{k^2}{\sqrt{\omega_k}}(\adk +a_k)+H_{vibr}$. Restricting
the application of this Hamiltonian to those TLSs such that
$|\Doz|\leq\Dox$ seems to be a fairly good approximation, but a
price has been paid, namely the spectral weight at $\om=0$ due to
the bias has been lost in this effectively symmetric spin-boson
approximation. This weight cannot be ignored, and it has to be added
as a different mechanism (the relaxational mechanism), what will be
done in the next subsection. Once this issue has been taken care
off, all the dissipative processes due to the presence of TLSs are
correctly included in this framework, and we can proceed describing
non-relaxational friction due to non-resonant, weakly damped, weakly
biased TLSs.

As mentioned before, the coupled system TLSs + vibrations can be
viewed, taking the coupling as a perturbation, from the point of
view of the excited mode $\wo$ as a set of TLSs with a modified
absorption spectrum. The TLSs, dressed perturbatively by the modes,
are entities capable of absorbing and emitting over a broad range of
frequencies, transferring energy from the excited mode $\wo$ to
other modes. The contribution to the value of the inverse of the
quality factor, $Q^{-1}(\wo)$, of all these non-resonant TLSs will
be proportional to
$A_{\rm{off-res}}^{\rm{tot}}(\wo)=\sum_{\Dox=0}^{\wo-\Gamma(\wo)}A(\Dox,\Doz,\wo)+
\sum_{\wo+\Gamma(\wo)}^{\e_{max}}A(\Dox,\Doz,\wo)\approx
2P\alf\sqrt{\wco/\wo}$ , a quantity measuring the density of states
which can be excited through $H_{\rm{int}}$ at frequency $\wo$, see
Appendix \ref{Aoffresonance} for details.

For an excited mode $\wo$ populated with $n_{\wo}$ phonons,
$Q^{-1}(\wo)$ is given by $Q^{-1}(\wo)=\Delta E / 2\pi E_0$, where
$E_0$ is the energy stored in the mode per unit volume, $E_0\simeq
n_{\wo}\hbar\wo / twL$, and $\Delta E$ is the energy fluctuations
per cycle and unit volume. $\Delta E$ can be obtained from Fermi
Golden Rule:
\begin{equation}\label{FGRrate}
    \Delta E_{\rm{off-res}}^{\rm{tot}}\simeq \frac{2\pi}{\wo}\times\hbar\wo\times\frac{2\pi}{\hbar}n_{\wo}
    \left(\lambda \frac{k_{0}^{2}}{\sqrt{\wo}}\right)^2 A_{\rm{off-res}}^{\rm{tot}}(\wo)\,,
\end{equation}
and the inverse quality factor of the vibration follows. For finite
temperatures the calculation of $A_{\rm{off-res}}^{\rm{tot}}(\wo,T)$
is done in Appendix \ref{Aoffresonance}. The result, valid for
temperatures below the breakdown of the TLS approximate description
of the two-well potential ($T\sim5$K$\,\,$\cite{E98}), is different
from the one given in $\,\,$\cite{SGN07}, because there the value of
$Q^{-1}(\wo,T)$ was interpreted as corresponding to the net energy
loss of the studied mode (subtracting emission processes from
absorption ones), while in experiments the \emph{observed linewidth}
is due to the total amount of \emph{fluctuations}, the addition of
emission and absorption processes. So in this context dissipation
means fluctuations, and not net loss of energy. The contribution of
these kind of processes is thus
\begin{equation}\label{Mainresult}
 (Q^{-1})_{\rm{off-res}}^{\rm{tot}}(\omega_0,T) \simeq 10 P_0t^{3/2} w\left(\frac{E}{\rho}\right)^{1/4} \frac{\alf^2\wco}{\wo}
 \rm{cotanh}\Bigl[\frac{\wo}{T}\Bigr]
\end{equation}

\subsection{Contribution of biased TLSs to the linewidth: relaxation absorption}
This very general friction mechanism arises due to the phase delay
between stress and imposed strain rate. In our context, for a given
TLS the populations of its levels take a finite time to readjust
when a perturbation changes the energy difference between its
eigenstates $\,\,$\cite{J72,E98}. This time $\tau$ corresponds to
the inverse linewidth and is given, for not too strong
perturbations, by the Fermi Golden Rule result
\begin{equation}\label{gammaresonant}
    \Gamma(\varepsilon,T)=16\alpha_b\sqrt{\wco}\times\sqrt{\varepsilon}\coth[\varepsilon/2T]
\end{equation}
where $\varepsilon=\sqrt{(\Dox)^2+(\Delta^z)^{2}}$ is the energy
difference between the levels. Notice that $\Delta^z$ is not just
the bare $\Doz$ appearing in the Hamiltonian of the system, but the
net bias including the modification due to the coupling to the
vibrational modes, $\Delta^z=\Doz+\xi_k
\partial u_{k}$ ($\xi_k$ is the corresponding coupling constant with
the proper dimensions and $\partial u_{k}$ a component $k$ of the
deformation gradient matrix, defined recalling
eq.(\ref{interaction}) as $\partial u_k\sim\langle k,
n_k|(k/\sqrt{\om_k})(\adk+a_k)|k, n_k\rangle$, associated to a
vibrational mode $|k,n_k\rangle$). Eq.(\ref{gammaresonant}) is valid
in the range of applicability of the TLS description of the two well
potentials ($T\leq5$K), and for values of $\alpha_b$ such that
$\Gamma(\e,T)<\e$. Therefore the energy levels
$\e_{1,2}=\mp\frac{1}{2}\sqrt{(\Dox)^2+(\Delta^z)^2}=\mp\frac{1}{2}\sqrt{(\Dox)^2+(\Doz+\xi_k
\partial u_{k})^2}$
depend on $\partial u_{k}$, and to first order the sensitivity of
these energies to an applied strain is proportional to the bias
\begin{equation}\label{Eofgamma}
    \frac{\partial \e_{i}}{\partial (\partial u_{k})}=\frac{(\Doz\mp\xi_k \partial u_{k})}{\e_{i}}\xi_k
    =\frac{\Delta_{z}^{tot}}{\e_{i}}\xi_k\,\,\,,
\end{equation}
with a response of the TLS $\propto (\Delta^{z}/\e)^2$. In
$\,\,$\cite{E98} a detailed derivation is given, and the imaginary
part of the response, corresponding to $Q^{-1}$ is also $\propto
\tau/(1+\om^2\tau^2)$, which in terms of $A(\om)$ is the lorentzian
peak at $\om=0$ of fig(\ref{spectrum}b).

\begin{figure}
\begin{center}
\includegraphics[width=7cm]{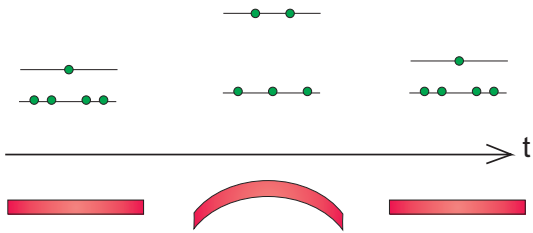}\\
\caption[fig]{\label{relaxation_delay} (Color online) Schematic
representation of the levels' population evolution of an ensemble of
5 identical TLSs with a response time $\tau\sim1/\om$, where $\om$
is the frequency of the bending mode excited, whose evolution is
also depicted in the lower part of the figure. The delay $\tau$ plus
the bias $\Doz$ give rise to the relaxational energy loss mechanism
of the mode, see text.}
\end{center}
\end{figure}

The mechanism is most effective when $\tau\sim1/\om$, where $\om$ is
the frequency of the vibrational mode; then, along a cycle of
vibration, the following happens (see fig.(\ref{relaxation_delay})):
When the TLS is under no stress the populations, due to the delay
$\tau$ in their response, are being still adjusted as if the levels
corresponded to a situation with maximum strain (and therefore of
maximum energy separation between them, cf. $\e_{1,2}(u_k)$), so
that the lower level becomes overpopulated. As the strain is
increased to its maximum value the populations are still adjusting
as if the levels corresponded to a situation with minimum strain,
thus overpopulating the upper energy level. Therefore in each cycle
there is a net absorption of energy from the mechanical energy
pumped into the vibrational mode.

If on the other hand $\tau\gg1/\om$ the TLSs levels' populations are
frozen with respect to that fast perturbation, while in the opposite
limit $\tau\ll1/\om$ the levels' populations are always in
instantaneous equilibrium with the variations of $\e_{1,2}$ and
there is neither a net absorption of energy.

For an ensemble of TLSs, the contribution to $Q^{-1}$ is
$\,\,$\cite{E98}
\begin{eqnarray}\label{QEsq}
    Q^{-1}_{\rm{rel}}(\omega,T)&=& P_0\gamma^2/(E T)
    \smallint_{0}^{\e_{max}}d\e
\smallint_{u_{min}}^{1} du \sqrt{1-u^2}/u \times \nonumber \\
 &\times & \frac{1}{\cosh^{2}\left(\e/2T\right)}\frac{\omega\tau}{1+(\omega\tau)^2}
\end{eqnarray}

The derivation of eq.(\ref{QEsq}) only relies in the assumptions of
an existence of well defined levels who need a finite time $\tau$ to
reach thermal equilibrium when a perturbation is applied, and the
existence of bias $|\Doz|>0$. This implies that such a scheme is
applicable also to our 1D vibrations, but is valid only if the
perturbation induced by the bath on the TLSs is weak, so that the
energy levels are still well defined. Therefore we will limit the
ensemble to underdamped TLSs for whom $\Gamma(\e,T)<\e$. In
eq.(\ref{QEsq}) the factor $\cosh^{-2}\left(\e/2T\right)$ imposes an
effective cutoff $\e<T$, so that in eq.(\ref{gammaresonant}) one can
approximate $\coth[\e/2T]\sim2T/\e$, resulting in
$\Gamma(\e)\sim1/\sqrt{\e}$, and thus the underdamped TLSs will
satisfy $\e \geq [30 \alf\sqrt{ \wco} T]^{2/3}$. For
$T\gg[32\alf\sqrt{\wco}]^2$, which is fulfilled for typical sizes
and temperatures (see Appendix \ref{Relanalysis} for details)
\begin{equation}\label{QEsq2}
Q^{-1}_{\rm{rel}}(\wo,T)\approx\frac{20P_0\gamma^4}{t^{3/2}w}\frac{(1+\nu)(1-2\nu)}{E^2(3-5\nu)}\Bigl(\frac{\rho}{E}\Bigr)^{1/4}\frac{\sqrt{T}}{\wo}
\end{equation}
Here $V_{amorph}\sim V_{tot}/10$ was assumed.

\section{Comparison between contributions to $Q^{-1}$. Relaxation prevalence.}
It is useful to estimate the relative importance of the
contributions to $Q^{-1}$ coming from the last two mechanisms. For
that sake, we particularize the comparison to the case of the
fundamental flexural mode, which is the one usually excited and
studied, of a doubly clamped beam, with frequency
$\wo\approx6.5(E/\rho)^{1/2}t/L^2$ (for a cantilever these
considerations hold, with only a slight modification of the
numerical prefactors; the conclusions are the same). The result is:
\begin{equation}\label{Comparison1}
 \Bigl[\frac{Q^{-1}_{\rm{rel}}(T)}{Q^{-1}_{\rm{off-res}}(T)}\Bigr]_{\rm{fund}}\approx\frac{300t^{1/2}}{L}
 \frac{(3-5\nu)}{(1+\nu)(1-2\nu)}\frac{E}{\rho}\frac{1}{T^{1/2}}
\end{equation}
For a temperature $T=1$K, the result is as big as $10^6$ even for a
favorable case $t=1$nm, $L=1\mu$m, $E=50$GPa, $\nu=0.2$,
$\rho=3$g/cm$^3$. So for any reasonable temperature and dimensions
the dissipation is dominated by the relaxation mechanism, so that
the prediction of the limit that surfaces set on the quality factor
of nanoresonators is, within this model for the surface,
eq.(\ref{QEsq2}):
\begin{equation}\label{QEsq3}
 Q^{-1}_{\rm{surface}}(\wo,T)\approx Q^{-1}_{\rm{rel}}(\wo,T)\sim\frac{T^{1/2}}{\wo}
\end{equation}

\begin{figure}
\begin{center}
\includegraphics[width=7cm]{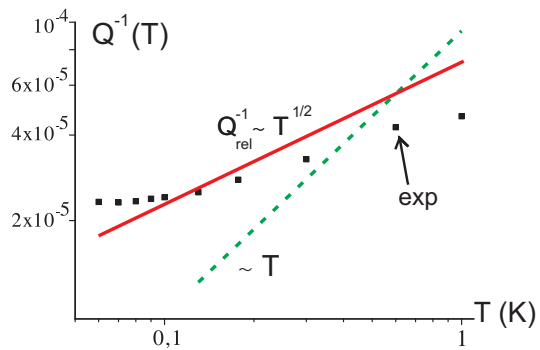}\\
\caption[fig]{\label{Comp1} (Color online) Example of fit of
$Q^{-1}_{\rm{rel}}(\wo,T)$, eq.(\ref{QEsq2}), to experimental data.
The data correspond to a Si resonator vibrating at $\wo=12.028$ MHz,
see ref.$\,$\cite{ZGSBM05}. Although the predicted order of
magnitude for the dissipation and the sublinear temperature
dependence is observed, the experimental trend follows a weaker
temperature dependence than the prediction $T^{1/2}$. For comparison
a linear temperature dependence is shown (dotted line). See the text
for more details.}
\end{center}
\end{figure}

For typical values $L\sim1\mu$m, $t,w\sim0.1\mu$m, $\gamma\sim5$ eV,
$P_0V_{amorph}/V_{tot}\sim10^{44}$J/m$^3$, and T in the range
1mK-0.5K the estimate for $Q^{-1}_{\rm{surface}}\sim10^{-4}$ gives
the observed order of magnitude in experiments like
$\,\,$\cite{ZGSBM05}, and also predicts correctly a sublinear
dependence, but with a higher exponent, $1/2$ versus the
experimental fit $0.36$ in $\,\,$\cite{ZGSBM05} or $0.32$ in
$\,\,$\cite{Mohetal07}, see fig.(\ref{Comp1}) for an example.

\section{Extensions to other devices}
\subsection{Cantilevers, nanopillars and torsional oscillators}
The extrapolation from doubly clamped beams to cantilevers
$\,\,$\cite{Metal05} and nanopillars $\,\,$\cite{SB04} is immediate,
the only difference between them being the allowed (k,$\om$(k))
values due to the different boundary conditions at the free end (and
even this difference disappears as one considers high frequency
modes, where in both cases one has $k_n\approx(2n+1)\pi/2L$). All
previous results apply, and one has just to take care in the
expressions corresponding to the $Q^{-1}$ of the fundamental mode,
where there is more difference between the frequencies of both
cases, the cantilever one being
$\wo^{cant}\approx(E/\rho)^{1/2}t/L^2$ as compared to the doubly
clamped case, $\wo^{clamped}\approx6.5(E/\rho)^{1/2}t/L^2$.
\subsection{Effect of the flexural modes on the dissipation of torsional
  modes}
The contribution from the TLSs + subohmic bending mode environment
to the dissipation of a torsional mode of a given oscillator can be
also estimated. We will study the easiest (and experimentally
relevant$\,\,$\cite{SRCR04}) case of a cantilever. For paddle and
double paddle oscillators the geometry is more involved, modifying
the moment of inertia and other quantities. When these changes are
included, the analysis follows the same steps we will show.

\textbf{Relaxation absorption.} We assume, based on the previous
considerations on the predominant influence of the flexural modes on
the TLSs dynamics, as compared with the influence of the other
modes, that the lifetime $\tau=\Gamma^{-1}$ of the TLSs is given by
eq.(\ref{gammaresonant}). The change in the derivation of the
expression for $Q^{-1}$ comes in eq.(\ref{Eofgamma}), where the
coupling constant $\xi_k^{\rm{tors}}$ is different, which translates
simply, in eq.(\ref{QEsq}), into substituting
$\gamma^2\leftrightarrow\gamma_{\rm{tors}}^2$, and the corresponding
prediction for $Q^{-1}_{\rm{rel}}$
\begin{equation}\label{QEsqtors}
Q^{-1}_{\rm{rel}}(\wo,T)\approx\frac{20P_0\gamma_{\rm{tors}}^2\gamma^2}{t^{3/2}w}\frac{(1+\nu)(1-2\nu)}{E^2(3-5\nu)}
\Bigl(\frac{\rho}{E}\Bigr)^{1/4}\frac{\sqrt{T}}{\wo}\,,
\end{equation}
where now $\wo$ is the frequency of the corresponding flexural mode.
The range of temperatures and sizes for which this result applies is
the same as in the case of an excited bending mode.

\textbf{Dissipation of symmetric non-resonant TLSs} The modified
excitation spectrum of the TLS's ensemble,
$A_{\rm{off-res}}^{\rm{tot}}(\wo,T)$, remains the same, and the
change happens in the matrix element of the transition probability
of a mode $|k_0 , n_0\rangle$ appearing in eq.(\ref{FGRrate}),
$(\lambda k_{0}^{2}/\sqrt{\wo})^2$. The operator yielding the
coupling of the bath to the torsional mode which causes its
attenuation is the interaction term of the Hamiltonian, which for
twisting modes is (see Appendix \ref{Tors} for the derivation of
eqs.(\ref{Hinttors})-(\ref{E0tors}))
\begin{equation}\label{Hinttors}
H_{int}^{tors}=\hbar \sigma_z \sum_k \gamma \sqrt{\frac{C}{8\mu tw}}
    \sqrt{\frac{1}{2\rho I \hbar L}}
    \frac{k}{\sqrt{\om_k}}(\adk +a_k)\,,
\end{equation}
where $\mu=E/[2(1+\nu)]$ is a Lande coefficient, and $C=\mu t^3w/3$
is the torsional rigidity. Again, $Q^{-1}(\om_j)=\Delta E / 2\pi
E_0$, where the energy $E_0$ stored in a torsional mode
$\phi_j(z,t)=A\sin[(2j-1)\pi z/(2L)]\sin(\om_j t)$ per unit volume
is $E_0=A^2\om_j^2\rho (t^2+w^2)/48$ ($z$ is the coordinate along
the main axis of the rod). Expressing the amplitude $A$ in terms of
phonon number $n_j$, the energy stored in mode $|k_j , n_j\rangle$
is
\begin{equation}\label{E0tors}
    E_0(k_j,n)=\frac{1}{2}\frac{\hbar\om_j}{(t^3w+w^3t)L}(t^2+w^2)(2n_j+1)\,,
\end{equation}
the energy fluctuations in a cycle of such a mode is
\begin{equation}\label{DeltaEtors}
\Delta E
=\frac{2\pi}{\om_j}\times\hbar\om_j\times\frac{2\pi}{\hbar}\frac{\gamma^2\hbar\om_j}{16Ltw\mu}n_jA_{\rm{off-res}}^{\rm{tot}}(\om_j,T)\,,
\end{equation}
and the inverse quality factor
\begin{eqnarray}\label{Qofftors}
 \nonumber(Q^{-1})_{\rm{off-res}}^{\rm{tot}}(\wo,T)&\approx& 0.04\frac{\gamma^4P}{t^{3/2}w}
  \frac{\rho^{1/4}(1+\nu)^2(1-2\nu)}{E^{9/4}(3-5\nu)}\times\\
  &&\times\frac{1}{\sqrt{\wo}}\rm{cotanh}\Bigl[\frac{\wo}{T}\Bigr]\,,
\end{eqnarray}
For sizes and temperatures as the ones used for previous estimates
the relaxation contribution dominates dissipation.

\section{Frequency shift}
Once the quality factor is known the relative frequency shift can be
obtained via a Kramers-Kronig relation (valid in the linear regime),
because both are related to the imaginary and real part,
respectively, of the acoustic susceptibility. First we will
demonstrate this, and afterwards expressions for beam and cantilever
will be derived and compared to experiments.
\subsection{Relation to the acoustic susceptibility}
In absence of sources of dissipation, the equation for the bending
modes is given by $-(12\rho/t^2)\partial^2 X/\partial
t^2=E\partial^4 X/\partial z^4$. The generalization in presence of
friction is
\begin{equation}\label{bendfriction}
    -\frac{12\rho}{t^2}\frac{\partial^2 X}{\partial t^2}=
(E+\chi)\frac{\partial^4 X}{\partial z^4}
\end{equation}
Where $\chi$ is a complex-valued susceptibility. Inserting a
solution of the form $X(z,t)=Ae^{i(kx-\om t)}$, where k is now a
complex number, one gets the dispersion relation
$\om=\sqrt{t^2(E+\chi)/(12\rho)}k^2$. Now, assuming that the
relative shift and dissipation are small, implying
$\rm{Re}(\chi)<<E$ , $\rm{Im}(k)<<\rm{Re}(k)$, the following
expressions for the frequency shift and inverse quality factor are
obtained in terms of $\chi$:
\begin{equation}\label{3dfreqshift}
    \left\{
      \begin{array}{ll}
        Q^{-1}&=\Delta\om/\om = -\rm{Im}(\chi)/E \\
        \delta\om/\om&=\rm{Re}(\chi)/2E
      \end{array}
    \right.
\end{equation}
Therefore a Kramers-Kronig relation for the susceptibility can be
used to obtain the relative frequency shift:
\begin{equation}\label{KK}
    \frac{\delta\om}{\om}(\om,T)=-\frac{1}{2\pi}P
\int_{-\infty}^{\infty}d\om'\frac{Q^{-1}(\om',T)}{\om'-\om}\,\,,
\end{equation}
where $P$ means here the principal value of the integral.

\subsection{Expressions for the frequency shift}
Relaxation processes of biased, underdamped TLSs dominate the
perturbations of the ideal response of the resonator, as we have
already shown for the inverse quality factor. For most of the
frequency range, $\om\geq [30 \alf\sqrt{ \wco} T]^{2/3}$,
$Q^{-1}(\om,T)\approx A\sqrt{T}/\om$ , with $A$ defined by
eq.(\ref{QEsq2}). The associated predicted contribution to the
frequency shift, using eq.(\ref{KK}), is
\begin{equation}\label{freqshift1}
    \frac{\delta\om}{\om}(\om,T)\approx-\frac{A}{2\pi}\frac{\sqrt{T}}{\om}\log\Bigl[\Bigl|1-\frac{\om}{[30 \alf\sqrt{ \wco} T]^{2/3}}\Bigr|\Bigr]
\end{equation}
For low temperatures, $\om\gg [30 \alf\sqrt{ \wco} T]^{2/3}$, the
negative shift grows towards zero as $\delta\om/\om
(\om,T)\sim\sqrt{T}\log[T^{2/3}/\om]$, reaching at some point a
maximum value, and decreasing for high temperatures, $\om< [30
\alf\sqrt{ \wco} T]^{2/3}$, as $\delta\om/\om (\om,T)\sim
1/T^{1/6}$. Even though the prediction of a peak in $\delta\om/\om
(T)$ qualitatively matches the few experimental results currently
available $\,\,$\cite{ZGSBM05,Mohetal07}, it does not fit with them
quantitatively.

\section{Applicability and further extensions of the model. Discussion}
As mentioned, the predictions obtained within this theoretical
framework do match qualitatively experimental results in terms of
observed orders of magnitude for $Q^{-1}(T)$, weak sublinear
temperature dependence, and presence of a peak in the frequency
shift temperature dependence. But quantitative fitting is still to
be reached, while on the experimental side more experiments need to
be done at low temperatures to confirm the, until now, few
results$\,\,$\cite{ZGSBM05}.

\textbf{Applicability.} The several simplifications involved in the
model put certain constraints, some of which are susceptible of
improvement. We enumerate them first and discuss some of them
afterwards: i) The probability distribution $P(\Dox,\Doz)$, borrowed
from amorphous bulk systems, may be different for the case of the
resonator's surface, ii) The assumption of non-interacting TLSs,
only coupled among them in an indirect way through their coupling to
the vibrations, breaks down at low enough temperatures, where also
the discreteness of the vibrational spectrum affects our predictions
iii) When temperatures rise above a certain value, high energy
phonons with 3D character dominate dissipation, the two-state
description of the degrees of freedom coupled to the vibrations is
not a good approximation, and thermoelastic losses begin to play an
important role, iv) For strong driving, anharmonic coupling among
modes has to be considered, and some steps in the derivation of the
different mechanisms, which assumed small perturbations, must be
modified. This will be the case of resonators driven to the
nonlinear regime, where bistability and other phenomena take place.

The solution to issue i) is intimately related to a better knowledge
of the surface and the different physical processes taking place
there. Recent studies try to shed some light on this
question$\,\,$\cite{C07}, and from their results a more realistic
$P(\Dox,\Doz)$ could be derived, which remains for future work.
Before that point, it is easier to wonder about the consequences of
a dominant kind of dissipative process which corresponded to a set
of TLSs with a well defined value of $\Dox$ and a narrow
distribution of $\Doz$'s of width $\Delta_1$, as was suggested for
single-crystal silicon$\,\,$\cite{P88}. Following $\,\,$\cite{P88},
a $Q^{-1}(T)\sim\sqrt{T}$ behavior is obtained for low temperatures
$T<\Delta_1$ if $\Gamma(\Dox,T)<\om$, and a
$Q^{-1}(T)\sim1/\sqrt{T}$ if $\Gamma(\Dox,T)>\om$, while at high
temperatures $T>\Delta_1$ a constant $Q^{-1}(T)\sim Q_0$ is
predicted for both cases. These predictions do not match better with
experiments than the results obtained with $P(\Dox,\Doz)\sim
P_0/\Dox$, so issue i) remains open. From the point of view of those
who attribute the origin of the low energy TLSs to the long range
interaction between localized defects, a possible source of change
for $P(\Dox,\Doz)$ may be the decrease in size of the resonator
below the correlation length of this interaction.

We try now to have a first estimate of the temperature for which
interactions between TLSs cannot be ignored. Following the ideas
presented in $\,\,$\cite{E98}, we will estimate the temperature $T*$
at which the dephasing time $\tau_{\rm{int}}$ due to interactions is
equal to the lifetime $\tau (T)=\Gamma^{-1}(T)$ defined in
eq.(\ref{gammaresonant}), for the TLSs that contribute most to
dissipation, which are those with
$\e=\sqrt{(\Dox)^2+(\Doz)^{2}}\sim\Dox\sim T$. For them
$\tau^{-1}(T)=\Gamma(T)\approx 40\alf\sqrt{\wco}\sqrt{T}$. The
interactions between the TLSs are dipolar, described by
$H_{\rm{int}}=\sum_{i,j}U_{1,2}\sigma_1^z\sigma_2^z$, with
$U_{1,2}=b_{12}/r_{12}^3$, $b_{12}$ verifying $\langle
b_{12}\rangle\approx0$, $\langle|b_{12}|\rangle\equiv
U_0\approx\gamma^2/E$ $\,\,$\cite{YL88,E98}. From the point of view
of a given TLS the interaction affects its bias,
$\Delta_j^z=(\Doz)_j+\sum_iU_{ij}\sigma_j^z$, causing fluctuations
of its phase $\delta \e_j(t)$ (where $t$ here means time and not
thickness), which have an associated $\tau_{\rm{int}}$ defined by
$\delta \e_j(\tau_{\rm{int}})\tau_{\rm{int}}\sim1$. These
fluctuations are caused by those TLSs which, within the time
$\tau_{\rm{int}}$, have undergone a transition between their two
eigenstates, affecting through the interaction $H_{\rm{int}}$ the
value of the bias of our TLS. At a temperature $T$, the most
fluctuating TLSs are those such that
$\e=\sqrt{(\Dox)^2+(\Doz)^{2}}\sim\Dox\sim T$, and their density can
be estimated, using $P(\Dox,\Doz)$, as $n_T\approx P_0kT$ they will
fluctuate with a characteristic time
$\tau(T)\approx[40\alf\sqrt{\wco}\sqrt{T}]^{-1}$, so for a time
$t<\tau(T)$ the amount of these TLSs that have made a transition is
roughly $n(t)\approx P_0kTt/\tau(T)$. For a dipolar interaction like
the one described above, the average energy shift is related to
$n(t)$ by $\,\,$\cite{BH77} $\delta\e(t)\approx U_0 n(t)$.
Substituting it in the equation defining $\tau_{\rm{int}}$, and
imposing $\tau_{\rm{int}}(T*)=\tau(T*)$ gives the transition
temperature $T*\approx[6\alf\sqrt{\wco}/(U_0P_0)]^2$. For example,
for a resonator like the silicon ones studied in
$\,\,$\cite{ZGSBM05}, $L=6\mu$m, $t=0.2\mu$m, $w=0.3\mu$m, the
estimated onset of interactions is at $T*\approx10$mK.

An upper limit $T_{\rm{high}}$ of applicability of the model due to
high energy 3D vibrational modes playing a significant role can be
easily derived by imposing $T_{\rm{high}}=\om_{\rm{min}}^{\rm{3D}}$.
The frequency $\om_{\rm{min}}^{\rm{3D}}$ corresponds to phonons with
wavelength comparable to the thickness $t$ of the sample (do not
confuse with time), $\om_{\rm{min}}^{\rm{3D}}=2\pi\sqrt{E/\rho}/t$.
The condition is very weak, as the value for example for silicon
resonators reads $T_{\rm{high}}\approx400/t$, with $t$ given in nm
and $T_{\rm{high}}$ in K. At much lower temperatures the two-state
description of the degrees of freedom coupled to the vibrations
ceases to be realistic, with a high temperature cutoff in the case
of the model applied to amorphous bulk systems of $T\sim5$K.

\begin{figure}
\begin{center}
\includegraphics[width=7cm]{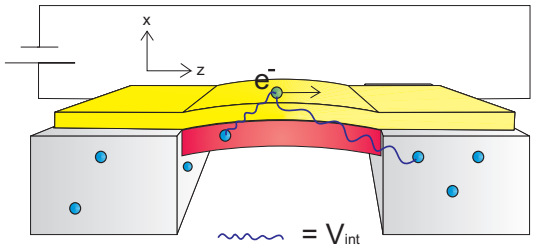}
    \caption{(Color online)Sketch of the distribution of charges in the device. When the
      system oscillates, these charges induce time dependent potentials which
      create electron-hole pairs in the metallic layer deposited on top of the beam, absorbing part of the mechanical energy of the flexural mode. See
      text for details.}
    \label{sketch_beam_charge}
\end{center}
\end{figure}
\section{Dissipation in a metallic conductor}
Many of the current realizations of nanomechanical devices monitor
the system by means of currents applied through metallic conductors
attached to the oscillators. The vibrations of the device couple to
the electrons in the metallic part. This coupling is useful in order
to drive and measure the oscillations, but it can also be a source
of dissipation. We will apply here the techniques described in
$\,\,$\cite{GJS04,G05} (see also $\,\,$\cite{WR95}) in order to
analyze the energy loss processes due to the excitations in the
conductor.

We assume that the leading perturbation acting on the electrons in
the metal are offset charges randomly distributed throughout the
device. A charge $q$ at position ${\bf \vec{R}}$ interacts with an
energy
\begin{equation}
V ( {\bf \vec{r}} , t ) \equiv \frac{q^2}{\epsilon_0 | {\bf \vec{R}}
( t ) -  {\bf \vec{r}} ( t ) |}
\end{equation}
with another charge $q$ at a position ${\bf \vec{r}}$ inside the
metal, see fig.[\ref{sketch_beam_charge}]. As the bulk of the device
is an insulator, this potential is only screened by a finite
dielectric constant, $\epsilon_0$. The oscillations of the system at
frequency $\wo$ modulate the relative distance $| {\bf \vec{R}} ( t
) - {\bf
  \vec{r}} ( t ) |$, leading to a time dependent potential acting on the
electrons of the metal.

The probability per unit time of absorbing a quantum of energy $\wo$
can be written, using second order perturbation theory, as
$\,\,$\cite{GJS04,G05}:
\begin{equation}
\Gamma = \int d {\bf \vec{r}}   d {\bf \vec{r}}' d t d t' V( {\bf
\vec{r}} , t )  V ( {\bf \vec{r}}' , t' )  \rm{Im} \chi [ {\bf
\vec{r}} - {\bf
    \vec{r}}' , t - t' ] e^{i \wo ( t - t' )}
\label{gamma_metal}
\end{equation}
where $\rm{Im} \chi [ {\bf \vec{r}} - {\bf
    \vec{r}}' , t - t' ]$ is the  imaginary part of the response function of
    the metal.

\textbf{Charges in the oscillating part of the device.} We write the
relative distance as ${\bf \vec{R}} ( t ) - {\bf
  \vec{r}} ( t ) = {\bf \vec{R}}_0 - {\bf \vec{r}}_0 + \delta {\bf \vec{R}}
  ( t ) - \delta {\bf \vec{r}} ( t ) $ and expand $V({\bf \vec{r}},t)$, whose time-dependent part is approximately
\begin{equation}
V ( {\bf \vec{r}} , t ) \approx \frac{q^2 \left[  {\bf \vec{R}}_0 -
{\bf
      \vec{r}}_0 \right] \cdot \left[ \delta {\bf \vec{R}}  ( t ) - \delta {\bf \vec{r}}
      ( t ) \right]}{\epsilon_0 | {\bf \vec{R}}_0 -  {\bf \vec{r}}_0 |^3}
\end{equation}
For a flexural mode, we have that $\left[ {\bf \vec{R}}_0 - {\bf
\vec{r}}_0 \right] \cdot[\delta {\bf \vec{R}}( t ) - \delta {\bf
\vec{r}}( t )]$ has turned into $[X_0-x_0]\cdot [\delta X(t) -
\delta x(t)]\sim t\cdot A\cdot\sin(\omega t)$, where $t=t_{\rm
ins}+t_{\rm metal}$ is the thickness of the beam and $A$ is the
amplitude of vibration of the mode. Thus the average estimate for
this case for the correction of $V({\bf \vec{r}},t)$ is
\begin{equation}
    \delta V ( {\bf \vec{r}} , t ) \sim \frac{q^2}{\epsilon_0}\frac{t\cdot
A\cdot\sin(\omega t)}{L^3}\,\,,
\end{equation}
$L$ being the resonator's length.

The integral over the region occupied by the metal in
eq.(\ref{gamma_metal}) can be written as an integral over ${\bf
\vec{r}}_0$ and ${\bf
  \vec{r}}_0'$. The dielectric constant of a dirty metal in the Random Phase Approximation is $\,\,$\cite{NP99}:
\begin{equation}
{\rm Im} \chi ( {\bf \vec{r}} - {\bf \vec{r}}' , t - t' ) = \int
\frac{d\om d{\bf \vec{q}}}{(2\pi)^4}e^{i {\bf \vec{q}} (
  {\bf \vec{r}} - {\bf \vec{r}}' )} e^{i \om ( t - t' )} \frac{| \om |}{e^4 D \nu |
  {\bf \vec{q}} |^2}
\label{susc}
\end{equation}
where $e$ is the electronic charge, $D = \hbar v_{\rm F} l$ is the
diffusion constant, $ v_{\rm F}$ is the Fermi velocity, $l$ is the
mean free path, and $\nu \approx ( k_{\rm F} t_{\rm metal} )^2 / (
\hbar v_{\rm F} )$ is the one dimensional density of states.

Combining eq.(\ref{gamma_metal}) and eq.(\ref{susc}), and assuming
that the position of the charge, ${\bf \vec {R}}_0$ is in a generic
point inside the beam, and that the length scales are such that
$k_{\rm F}^{-1} , t_{\rm  metal} , t \ll L$, we can obtain the
leading dependence of $\Gamma$ in eq.(\ref{gamma_metal}) on $L$:
\begin{equation}
\Gamma \approx \frac{| \wo | A^2}{D \nu L} \left( \frac{t}{L}
\right)^2 \approx \frac{| \wo | A^2}{l k_{\rm F}^2 L^3} \left(
\frac{t}{t_{\rm metal}} \right)^2
\end{equation}
where we also assume that $| q | = e$. The energy absorbed per cycle
of oscillation and unit volume will be $\Delta
E=(2\pi/\wo)\hbar\wo\Gamma_{\rm ph}/t^2L=2\pi\hbar\Gamma_{\rm
ph}/t^2L$, and the inverse quality factor $Q^{-1}_{\rm ph}(\wo)$
will correspond to
\begin{equation}\label{Qfactor1a}
    Q^{-1}_{\rm ph}(\wo)=\frac{1}{2\pi}\frac{\Delta E}{E_0}=\frac{\hbar\Gamma_{\rm ph}}{twL}\frac{1}{\frac{1}{2}\rho\wo^2A^2}\,\,,
\end{equation}
where $E_0$ is the elastic energy stored in the vibration and $A$
the amplitude of vibration. Substituting the result for $\Gamma$ one
obtains
\begin{equation}
Q^{-1} \approx \frac{2 \hbar}{l k_{\rm F}^2 L^4t^2\rho\wo} \left(
\frac{t}{t_{\rm metal}} \right)^2 \label{Q_metal}
\end{equation}
In a narrow metallic wire of width $t_{\rm metal}$, we expect that
$l \sim t_{\rm metal}$.

Typical values for the parameters in eq.(\ref{Q_metal}) are $k_{\rm
F}^{-1} \approx 1$\AA, $A \approx 1$\AA, $l \sim t_{\rm metal}
\approx 10$nm $\sim 10^2$\AA, $t \approx 100$nm $\approx 10^3$\AA
and $L \approx 1 \mu$m $\approx 10^4$\AA. Hence, each charge in the
device gives a contribution to $Q^{-1}$ of order $10^{-20}$. The
effect of all charges is obtained by summing over all charges in the
beam. If their density is $n_q$, we obtain:
\begin{equation}
Q^{-1} \approx \frac{2 \hbar n_q}{l k_{\rm F}^2 L^3\rho\wo} \left(
\frac{t}{t_{\rm metal}} \right)^2 \label{Q_metal_n}
\end{equation}
For reasonable values of the density of charges, $n_q = l_q^{-3} ,
l_q \gtrsim 10$nm, this contribution is negligible, $Q^{-1} \lesssim
10^{-16}$.

\textbf{Charges in the substrate surrounding the device.}Many
resonators, however, are suspended, at distances much smaller than
$L$, over an insulating substrate, which can also contain unscreened
charges. As the Coulomb potential induced by these charges is long
range, the analysis described above can be applied to all charges
within a distance of order $L$ from the beam. Moreover, the motion
of these charges is not correlated with the vibrations of the beam,
so that now the value of $\left| \delta {\bf \vec{R}}  ( t ) -
\delta {\bf \vec{r}} ( t ) \right|$ has to be replaced by:
\begin{equation}
\left| \delta {\bf \vec{R}}  ( t ) - \delta {\bf \vec{r}} ( t )
\right| \approx A e^{i \wo t}\,\,, \label{rel_motion_2}
\end{equation}
and the value of $|{\bf \vec{R}}_0 - {\bf\vec{r}}_0|\sim L$.
Assuming, as before, a density of charges $n_q = l_q^3$, the effect
of all charges in the substrate leads to:
\begin{equation}\label{Qmetallic}
Q^{-1} \approx \frac{2 \hbar L}{l ( k_{\rm F} t_{\rm metal} )^2
l_q^3 t^2\rho\wo}\approx 0.3\frac{\hbar L^3}{l ( k_{\rm F} t_{\rm
metal} )^2 l_q^3 t^3\sqrt{E\rho}}\,\,,
\end{equation}
where the second result corresponds to the fundamental mode,
$\wo\approx 6.5(t/L^2)\sqrt{E/\rho}$. For values $L \approx 1 \mu$m,
$A \approx 1$\AA , $k_{\rm F}^{-1} \approx 1$\AA, $l \sim t_{\rm
metal} \approx 10$nm and $l_q \sim 10$nm, we obtain $Q^{-1} \sim
10^{-9}$. Thus, given the values of $Q^{-1}$ reached experimentally
until now this mechanism can be disregarded, although it sets a
limit to $Q^{-1}$ at the lowest temperatures. It also has to be
noted that this estimate neglects cancelation effects between
charges of opposite signs.
\section{Conclusions}
Disorder and configurational rearrangements of atoms and adsorbed
impurities at surfaces of nanoresonators dominate dissipation of
their vibrational eigenmodes at low temperatures. We have given a
theoretical framework to describe in a unified way these processes,
improving and extending previous ideas$\,\,$\cite{SGN07}. Based on
the good description of low temperature properties of disordered
bulk insulators provided by the Standard Tunneling Model
$\,\,$\cite{AHV72,P72,E98}, and in particular of acoustic phonon
attenuation in such systems, we adapt it to describe the damping of
1D flexural and torsional modes of NEMS associated to the
amorphous-like nature of their surfaces.

Correcting some aspects of $\,\,$\cite{SGN07}, we have calculated
the damping of the modes by the presence of an ensemble of
independent Two-Level Systems (TLSs) coupled to the local
deformation gradient field $\partial_iu_j$ created by vibrations.
The different dissipation channels to which this ensemble gives rise
have been described, focussing the attention on the two most
important: relaxation dynamics of biased TLSs and dissipation due to
symmetric non-resonant TLSs. The first one is caused by the finite
time it takes for the TLSs to readjust their equilibrium populations
when their bias $\Doz$ is modified by local strains, with biased
TLSs playing the main role, as this effect is
$\propto[\Doz/\sqrt{(\Doz)^2+(\Dox)^2}]^2$. In terms of the
excitation spectrum of the TLSs, it corresponds to a lorentzian peak
around $\om=0$. The second effect is due to the modified absorption
spectrum of the TLSs caused by their coupling to all the vibrations,
specially the flexural modes, whose high density of states at low
energies leads to subohmic damping$\,\,$\cite{Letal87,KM96,W99}. A
broad incoherent spectral strength is generated, enabling the
"dressed-by-the-modes" TLSs to absorb energy of an excited mode and
deliver it to the rest of the modes when they decay.

We have given analytical expressions for the contributions of these
mechanisms to the linewidth of flexural (eqs.(\ref{Mainresult}),
(\ref{QEsq2})) or torsional modes (eqs.(\ref{QEsqtors}),
(\ref{Qofftors})) in terms of the inverse quality factor
$Q^{-1}(\wo)=\Delta\wo/\wo$, showing the dependencies on the
dimensions, temperature and other relevant parameters characterizing
the device. We have compared the two mechanisms, concluding that
relaxation dominates dissipation, with a predicted
$Q^{-1}(\wo,T)\sim T^{1/2}/\wo$ . Expressions have been provided for
damping of flexural modes in cantilevers and doubly-clamped beams,
as well as for damping of their torsional modes.

Analytical predictions for associated frequency shifts have been
also calculated (eq.(\ref{freqshift1})). Some important successes
have been achieved, like the qualitative agreement with a sublinear
temperature dependence of $Q^{-1}(T)$, the presence of a peak in the
frequency shift temperature dependence $\delta\om/\om(T)$, or the
observed order of magnitude of $Q^{-1}(T)$ in the existing
experiments studying flexural phonon attenuation at low
temperatures$\,\,$\cite{ZGSBM05,Mohetal07}. Nevertheless, the lack
of full quantitative agreement has led to a discussion on the
assumptions of the model, its links with the physical processes
occurring at the surfaces of NEMS, its range of applicability and
improvements to reach the desired quantitative fit.

Finally, we have also considered the contributions to the
dissipation due to the presence of metallic electrodes deposited on
top of the resonators, which can couple to the electrostatic
potential induced by random charges. We have shown that the coupling
to charges within the vibrating parts does not contribute
appreciably to the dissipation. Coupling to charges in the
substrate, although more significant, still leads to small
dissipation effects (eq.(\ref{Qmetallic})), imposing a limit at low
temperatures $Q^{-1}\sim10^{-9}$, very small compared to the values
reached in current experiments $\,\,$\cite{ZGSBM05,Mohetal07}.

\section{Acknowledgements}
C. S. and F. G. acknowledge funding from MEC (Spain) through FPU
grant, grant FIS2005-05478-C02-01 and the Comunidad de Madrid,
through the program CITECNOMIK, CM2006-S-0505-ESP-0337. A.H.C.N. is
supported through NSF grant DMR-0343790.

\appendix
\section{Calculation of the spectral function for the
bending modes}\label{DerivJ} The starting point is the Hamiltonian
of ref.$\,\,$\cite{SGN07}, $\textsl{H}= \Delta_{0}\sigma_x + \gamma
\sigma_z \partial_i u_j$, where $\partial_i u_j$ is a component of
the deformation gradient matrix. In the case of the bending modes of
a rod of dimensions $L$, $t$ and $w$, and mass density $\rho$, there
are two variables $X(z),Y(z)$ (transversal displacements of the rod
as a function of the position along its length, z) obeying
$\,\,$\cite{LL59}
\begin{eqnarray}
    \nonumber EI_y\frac{\partial^4 X}{\partial z^4}&=&- \rho tw \frac{\partial^2 X}{\partial t^2}\\
    EI_x\frac{\partial^4 Y}{\partial z^4}&=&- \rho tw \frac{\partial^2 Y}{\partial t^2}
\end{eqnarray}
 (where $I_y=t^3 w/12$, and $I_x=w^3 t/12$), so that there are plane waves $X(z,t),Y(z,t)\sim e^{i(kz-\omega
t)}$, but with a quadratic dispersion relation, $\omega_{j}(k) =
\sqrt{\frac{EI_j}{\rho ab}}\times k^2$. One can thus express
$X(z),Y(z)$ in terms of bosonic operators, for example $X(0)=\sum_k
\sqrt{\frac{\hbar}{2\rho twL\om_k}}(a_{k}^{\dag}+a_k)$. We can
relate this variables to the strain field $\partial_i u_j$ through
the free energy F:
\begin{widetext}
\begin{equation}
    F_{rod}=\frac{1}{2}\int dz EI_y \Bigl ( \frac{\partial^2 X}{\partial z^2}  \Bigr )^2 +
   EI_x \Bigl ( \frac{\partial^2 Y}{\partial z^2}  \Bigr )^2=\frac{1}{2}\int dz \int dS \frac{1}{2}\lambda \sum_iu_{ii}^{2}+\mu \sum_{i,k}u_{ik}^{2}
   \approx \frac{1}{2}\int dz \int dS (\frac{3}{2}\lambda +9\mu)u_{ij}^{2}
\end{equation}
\end{widetext} extracting an average equivalence for one component
$u_{ij}$, $u_{ij}\approx2\sqrt{EI_y/(3\lambda + 18\mu)tw}\partial^2
X/\partial z^2$. The interaction term in the Hamiltonian is then
\begin{widetext}
\begin{equation}
    H_{int}=\hbar \sigma_z \sum_k\lambda\frac{k^2}{\sqrt{\omega_k}}(\adk +a_k)=\hbar \sigma_z
 \sum_{ij}\sum_k \Bigl [2\gamma \frac{\Delta_{0}^{x}}{\Delta_{0}}\sqrt{\frac{EI_y}{(3\lambda
 + 18\mu)tw}}\sqrt{\frac{1}{2\rho tw \hbar L}} \Bigr]\frac{(k^{ij})^2}{\sqrt{\omega_k^{ij}}}(a_{k}^{ij\dag}+a_{k}^{ij})
\end{equation}
\end{widetext} So we have approximately 9 times the same Hamiltonian,
once for each $u_{ij}$, and the corresponding spectral function
$J(\omega)$ will be nine times the one calculated for
\begin{widetext}
\begin{equation}\label{deflambda}
    H_{int}=\hbar \sigma_z \sum_k\lambda\frac{k^2}{\sqrt{\omega_k}}(\adk +a_k)\simeq\hbar \sigma_z\sum_k \Bigl [2\gamma
 \frac{\Delta_{0}^{x}}{\Delta_{0}}\sqrt{\frac{EI_y}{(3\lambda+ 18\mu)tw}}\sqrt{\frac{1}{2\rho tw \hbar L}}\Bigr
]\frac{k^2}{\sqrt{\omega_k}}(\adk +a_{k})
\end{equation}
\end{widetext}
For a Hamiltonian of the class $H = \Delta_{0}\sigma_x + \hbar
\sigma_z \sum_k \lambda_k(\adk +a_k)$ the spectral function
$J(\omega)$ is given by
$J(\omega)=\frac{1}{2\pi}\sum_k\lambda_{k}^{2} \delta (\omega -
\omega_k)$, so that in our case the expression for it is
\begin{widetext}
\begin{equation}
    J(\omega)=\frac{1}{2\pi}\sum_k \Bigl[2\gamma
 \frac{\Delta_{0}^{x}}{\Delta_{0}}\sqrt{\frac{EI_y}{(3\lambda+ 18\mu)tw}}\sqrt{\frac{1}{2\rho tw \hbar L}}
\frac{k^2}{\sqrt{\omega_k}} \Bigr]^2 \delta (\omega - \omega_k)
\end{equation}
\end{widetext}
Taking the continuum limit ($\frac{1}{L}\sum_k
\rightarrow \frac{1}{\pi}\int dk$):
\newpage
\begin{widetext}
\begin{equation}
    J(\omega)=\frac{2L}{(2\pi)^2}\int_{k_{min}}^{k_{max}} dk \Bigl[2\gamma
 \frac{\Delta_{0}^{x}}{\Delta_{0}}\sqrt{\frac{EI_y}{(3\lambda+ 18\mu)tw}}\sqrt{\frac{1}{2\rho tw \hbar L}}
\Bigr]^2\frac{k^4}{\omega_k} \delta (\omega - \omega_k)
\end{equation}
\end{widetext}
Using the dispersion relation $\omega_{j}(k) =
\sqrt{\frac{EI_j}{\rho ab}}\times k^2=c\times k^2$ we express the
integral in terms of the frequency:
\begin{widetext}
\begin{equation}
    J(\omega)=\frac{L}{(2\pi)^2}\int_{\omega_{min}}^{\wco} \frac{d\omega_k}{\sqrt{c\,\omega_k}} \Bigl[2\gamma
 \frac{\Delta_{0}^{x}}{\Delta_{0}}\sqrt{\frac{EI_y}{(3\lambda+ 18\mu)tw}}\sqrt{\frac{1}{2\rho tw \hbar L}}
\Bigr]^2\frac{k^4}{\omega_k} \delta (\omega -
\omega_k)=\frac{L}{(2\pi)^2}\Bigl[2\gamma
 \frac{\Delta_{0}^{x}}{\Delta_{0}}\sqrt{\frac{EI_y}{(3\lambda+ 18\mu)tw}}\sqrt{\frac{1}{2\rho tw \hbar L}}
\Bigr]^2\frac{\sqrt{\omega}}{c^{5/2}}
\end{equation}
\end{widetext}
$J_{{\rm flex}}(\omega)$ is just 9 times this, eq.(\ref{Jsubohmic}).

\section{Dissipation from off-resonance dressed
TLS$\rm{s}$}\label{Aoffresonance}

We follow the method of ref.$\,\,$\cite{G85}. The form of $A(\om)$,
the spectral function of a single TLS, for frequencies $\om \ll \D $
and $\om \gg \D $ can be estimated using perturbation theory.
Without the interaction, the ground state $|s\rangle$ of the TLS is
the symmetric combination of the ground states of the two wells, and
the excited state is the antisymmetric one, $|a\rangle$. We will use
Fermi's Golden Rule applied to the subohmic spin-boson Hamiltonian,
$H=\D \sigma_x +
    \hbar\lambda \sigma_z \sum_k \Bigl [ \frac{k^2}{\sqrt{\om_k}}\Bigr ](\adk +a_k)
    +\sum_k\hbar\om(k) a_{k}^{\dag}a_{k}$, where $a_k$ is the annihilation operator of a bending mode $k$.
Considering only the low energy modes $\om(k) \ll \D  $, to first
order the ground state and a state with energy $\om(k)$ are given by
\begin{eqnarray}
  \nonumber |g\rangle &\simeq& |s\rangle - \frac{\frac{\lambda k^2}{\sqrt{\om_k}}}{2\D  }
  a_{k}^{\dag}|a\rangle + ...\hphantom{xx},\\
  |k\rangle &\simeq& a_{k}^{\dag}|s\rangle - \frac{\frac{\lambda k^2}{\sqrt{\om_k}}}{2\D  }
  |a\rangle + ...\hphantom{xx}.
\end{eqnarray}
We estimate the behavior of $A(\om)$ by taking the matrix element of
$\sigma_z$ between these two states, obtaining (remember that
$\om(k)\propto k^2$)
\begin{equation}\label{deltabig}
    A(\om)\sim \frac{\hbar \alpha_b
    \sqrt{\wco}\sqrt{\om}}{\D  ^{2}}+ ...\hphantom{xx},\hphantom{xx}\om(k) \ll \D  .
\end{equation}
The expression in the numerator is proportional to the spectral
function of the coupling,  $J(\om)=\alpha_b \sqrt{\wco}\sqrt{\om}$.
Now we turn our attention to the case $\om(k) \gg \D  $, where the
ground state $|g \rangle$ and an excited state $|k\rangle$ can be
written as
\begin{eqnarray}
  \nonumber |g\rangle &\simeq& |s\rangle - \frac{\lambda k^2 /\sqrt{\om_k}}
  {\hbar \om_k + 2\D  }a_{k}^{\dag}|a\rangle + ...\hphantom{xx},\\
  |k\rangle &\simeq& a_{k}^{\dag}|s\rangle + \frac{\lambda k^2/\sqrt{\om_k}}
  {\hbar \om_k - 2\D  }|a\rangle + ...\hphantom{xx}.
\end{eqnarray}
The matrix element $\langle 0 |\sigma_z|k \rangle$ is $\sim
\frac{\lambda k^2 }{\sqrt{\om_k}}\frac{4\D  }{(\hbar \om_k)^2}$,
leading to
\begin{equation}\label{deltasmall}
    A(\om)\sim \frac{\alpha_b
    \sqrt{\wco}\D  ^{2}}{\hbar^3\om^{7/2}}+ ...\hphantom{xx},\hphantom{xx}\om(k) \gg \D  .
\end{equation}
\subsection{Value of $A_{\rm{off-res}}^{\rm{tot}}(\wo)$}
Now we will add the contributions of all the non-resonant TLSs using
the probability distribution $P(\Dox,\Doz)=P_0/\Dox$
$\,\,$\cite{AHV72,P72}. For the case of weak coupling, $\alf<1/2$
and $\wo \geq (2\alpha_b)^2\wco$, which is the one found in
experiments, one has
\begin{eqnarray}\label{withresonance}
      \nonumber A_{\rm{off}}^{\rm{tot}}(\wo) \sim \int_{\hbar[\wo+\Gamma(\wo)]}^{\e_{max}}&&d\Dox\int_{-\Dox}^{\Dox}
      d\Dz\frac{P}{\Dox}\frac{\hbar \alpha_b \sqrt{\wco}\sqrt{\wo}}{(\Dox)^{2}}\\
       \nonumber +\int_{\hbar(2\alpha_b)^2\wco}^{\hbar[\wo-\Gamma(\wo)]}
    &&d\Dox \int_{-\Dox}^{\Dox}d\Dz\frac{P}{\Dox}\frac{\alpha_b\sqrt{\wco}(\Dox)^{2}}{\hbar^3\wo^{7/2}}
    \end{eqnarray}
obtaining the result $A_{\rm{off-res}}^{\rm{tot}}(\wo)\approx 2 P
\alpha_b \sqrt{\wco / \wo}$.
\subsection{The off-resonance contribution for T $>$ 0}
Using the same scheme, the modifications due to the temperature will
appear in the density of states of absorption and emission of energy
corresponding to a "dressed" TLS, $A(\D,\om,T)$. Now there will be a
probability for the TLS to be initially in the excited antisymmetric
state, $|a\rangle$, proportional to $\exp[-\D/kT]$, and to emit
energy $\hbar\om$ giving it to our externally excited mode,
$|k,n\rangle$, thus compensating the absorption of energy
corresponding to the opposite case (transition from
$|s\rangle|k,n\rangle$ to $|a\rangle|k,n-1\rangle$), but
contributing in an additive manner to the total amount of
fluctuations, which are the ones defining the linewidth of the
vibrational mode observed in experiments, fixing the value of
$Q^{-1}(\om,T)$. The expression for $A(\D,\om,T)$ is given by
\begin{equation}\label{AofT}
    A(\om,T)=\frac{1}{Z}\sum_{i}\sum_{f}|\langle
    i|\sigma_{z}|f\rangle|^2 e^{-\frac{E_i}{kT}}\delta[\hbar\om-(E_f-E_i)]
\end{equation}
We consider a generic state $|i_a\rangle=|a\rangle|k_1 n_1,...,k_j
n_j,...\rangle$ or $|i_s\rangle=|s\rangle|k_1 n_1,...,k_j
n_j,...\rangle$, and states that differ from it in $\hbar\om_j$,
$|f_{a\pm}\rangle=|a\rangle|k_1 n_1,...,k_j n_j\pm1,...\rangle$ and
$|f_{s\pm}\rangle|s\rangle|k_1 n_1,...,k_j n_j\pm1,...\rangle$. As
 for $T = 0$, we will correct them to first order in the
interaction Hamiltonian $H_{int}= \hbar\lambda \sigma_z \sum_k
\sqrt{\om_k}(\adk +a_k)$, and then calculate the square of the
matrix element of $\sigma_z$, $|\langle i|\sigma_{z}|f\rangle|^2$.

Elements $|\langle i_{a,s}|\sigma_{z}|f_{a+,s+}\rangle|^2$
correspond to absorption by the "dressed" TLS of an energy
$\hbar\om_j$ from the mode $k_j$, while elements $|\langle
i_{a,s}|\sigma_{z}|f_{a-,s-}\rangle|^2$ correspond to emission and
"feeding" of the mode with a phonon $\hbar\om_j$. The expressions
for the initial states are
\begin{widetext}
\begin{eqnarray}
    \nonumber |i_s\rangle=|s\rangle|k_1 n_1,...,k_j n_j,...\rangle &\rightarrow& |s\rangle|k_1 n_1,...,k_j n_j,...\rangle +
  \sum_{k_i\ni n_i>0}\frac{\hbar\lambda\sqrt{n_i\om_i}}{\hbar\om_i+2\D  }|a\rangle|...k_i n_i-1...\rangle\\
  \nonumber &-& \sum_{\forall k_i}\frac{\hbar\lambda\sqrt{(n_i+1)\om_i}}{\hbar\om_i-2\D  }|a\rangle|...k_i n_i+1...\rangle \\
  \nonumber |i_a\rangle=|a\rangle|k_1 n_1,...,k_j n_j,...\rangle &\rightarrow& |a\rangle|k_1 n_1,...,k_j n_j,...\rangle +
  \sum_{k_i\ni n_i>0}\frac{\hbar\lambda\sqrt{n_i\om_i}}{\hbar\om_i-2\D  }|s\rangle|...k_i n_i-1...\rangle\\
  &-& \sum_{\forall k_i}\frac{\hbar\lambda\sqrt{(n_i+1)\om_i}}{\hbar\om_i+2\D  }|s\rangle|...k_i n_i+1...\rangle
\end{eqnarray}
\end{widetext}
and, for example, the state $|f_{a+}\rangle$ is given by
\begin{widetext}
\begin{eqnarray}
  \nonumber |f_{a+}\rangle=|a\rangle|k_1 n_1,...,k_j n_j+1,...\rangle&\rightarrow& |a\rangle|k_1 n_1,...,k_j n_j+1,...\rangle +
  \sum_{k_i\ni n_i>0,i\neq j}\frac{\hbar\lambda\sqrt{n_i\om_i}}{\hbar\om_i-2\D  }|s\rangle|...k_i n_i-1...k_j n_j+1...\rangle \\
   \nonumber&+& \frac{\hbar\lambda\sqrt{(n_j+1)\om_j}}{\hbar\om_j-2\D  }|s\rangle|...k_j
   n_j...\rangle - \sum_{\forall k_i,i\neq j}\frac{\hbar\lambda\sqrt{(n_i+1)\om_i}}{\hbar\om_i+2\D  }|s\rangle|...k_i n_i+1...k_j
   n_j+1...\rangle\\
   &-&\frac{\hbar\lambda\sqrt{(n_j+2)\om_j}}{\hbar\om_j+2\D  }|s\rangle|...k_j n_j+2...\rangle
\end{eqnarray}
\end{widetext}
with similar expression for the rest of states. The
value of $|\langle i_{a,s}|\sigma_{z}|f_{a+,s+}\rangle|$
(absorption) is $|\langle
i_{a,s}|\sigma_{z}|f_{a+,s+}\rangle|=\Bigl|\frac{\lambda\sqrt{(n_j+1)\om_j}4\D
}{(\hbar\om_j)^2-4\D  ^2}\Bigr|$ , with $n_j=0,1,...$, while for
emission the result is $|\langle
i_{a,s}|\sigma_{z}|f_{a-,s-}\rangle|=\Bigl|\frac{\lambda\sqrt{n_j\om_j}4\D
}{(\hbar\om_j)^2-4\D  ^2}\Bigr|$,
 with $n_j=1,...$. Taking the limits we considered at T = 0
($\hbar\om_j\ll\D  $ and $\hbar\om_j\gg\D  $) the results are the
same as for T = 0 for absorption, except for a factor $(n_j+1)$, and
we also have now the possibility of emission, with the same matrix
element but with the factor $n_j$:
\begin{eqnarray}\label{matrixelT}
    \nonumber &\rm{Absorption}&\left\{
                       \begin{array}{ll}
                         \sim \frac{(n_j+1)\hbar\alf\sqrt{\wco}\sqrt{\om}}{\D  ^2} & \hbar\om_j\ll\D   \\
                         \sim \frac{(n_j+1)\alf\sqrt{\wco}\D  ^2}{\hbar^3\om^{7/2}}, & \hbar\om_j\gg\D
                       \end{array}
                     \right.\\
     &\rm{Emission}&\left\{
                       \begin{array}{ll}
                         \sim \frac{n_j\hbar\alf\sqrt{\wco}\sqrt{\om}}{\D  ^2} & \hbar\om_j\ll\D   \\
                         \sim \frac{n_j\alf\sqrt{\wco}\D  ^2}{\hbar^3\om^{7/2}}, & \hbar\om_j\gg\D
                       \end{array}
                     \right.
\end{eqnarray}
Now we have to sum over all initial states, and noting that the
first order correction to the energy of any eigenstate is 0, the
partition function, Z, is easy to calculate, everything factorizes,
and the result is, for example in the case $\hbar\om_j\ll\D  $:
\begin{widetext}
\begin{eqnarray}
  \nonumber &&A_{abs}(\D  ,\om_j,T) = \frac{1}{Z}\sum_{i\neq j,n_i=0}^{\infty}\sum_{n_j=0}^{\infty}
  \frac{\hbar\alf\sqrt{\wco}\sqrt{\om_j}}{\D  ^2}(n_j+1)e^{-\frac{\hbar\om_jn_j}{kT}}
  e^{-\frac{\hbar\sum_in_i\om_i}{kT}}=\frac{\hbar\alf\sqrt{\wco}}{\D  ^2}\frac{\sqrt{\om_j}}
  {1-e^{-\frac{\hbar\om_j}{kT}}}  \\
  && A_{em}(\D  ,-\om_j,T) = \frac{1}{Z}\sum_{i\neq j,n_i=0}^{\infty}\sum_{n_j=1}^{\infty}
  \frac{\hbar\alf\sqrt{\wco}\sqrt{\om_j}}{\D  ^2}n_je^{-\frac{\hbar\om_jn_j}{kT}}
  e^{-\frac{\hbar\sum_in_i\om_i}{kT}}=\frac{\hbar\alf\sqrt{\wco}}{\D  ^2}
  \frac{\sqrt{\om_j}e^{-\frac{\hbar\om_j}{kT}}}{1-e^{-\frac{\hbar\om_j}{kT}}}
\end{eqnarray}
\end{widetext}
In $A_{abs}(\D  ,\om_j,T)$ we have added the contributions from the
matrix elements $|\langle i_s|\sigma_{z}|f_{s+}\rangle|^2e^{\frac{\D
}{kT}}+|\langle i_a|\sigma_{z}|f_{a+}\rangle|^2e^{\frac{-\D
}{kT}}=|\langle i_s|\sigma_{z}|f_{s+}\rangle|^2[e^{\frac{\D
}{kT}}+e^{\frac{-\D }{kT}}]$ . The sum of exponentials cancels with
the partition function of the TLS (appearing as a factor in the
total Z), leading in this way to the expression above (the same
applies for $A_{em}(\D  ,-\om_j,T)$). The total fluctuations will be
proportional to their sum, which thus turns out to be at this level
of approximation $\propto\rm{cotanh}[\hbar\om_j/kT]$:
\begin{widetext}
\begin{equation}
A_{diss}(\D ,\om_j,T)=A_{abs}(\D ,\om_j,T)+A_{em}(\D
,-\om_j,T)=A_{diss}(\D
,\om_j,T=0)\rm{cotanh}\Bigl[\frac{\hbar\om_j}{kT}\Bigr]
\end{equation}
\end{widetext}
In fact, this result applies for any other type of modes,
independently of its dispersion relation, provided the coupling
Hamiltonian is linear in $\sigma_z$ and $(a_{-k}^{\dagger}+a_{k})$
and everything is treated at this level of perturbation theory.
Moreover, it can be proven that if one has an externally excited
mode with an average population $\langle n_j\rangle$ , and
fluctuations around that value are thermal-like, with a probability
$\propto \exp[-|n_j-\langle n_j\rangle|/kT)]$, one recovers again
the same temperature dependence,
$\propto\rm{cotanh}[\hbar\om_j/kT]$.

\section{Derivation of $Q^{-1}_{\rm{rel}}(\wo,T)$, eq.(\ref{QEsq2})}\label{Relanalysis}
As discussed after eq.(\ref{QEsq}), we have to sum over underdamped
TLSs, $\e \geq [30 \alf\sqrt{ \wco} T]^{2/3}$, using the
approximation for $\Gamma$,
$\Gamma(\e,T)\sim30\alf\sqrt{\wco}T/\sqrt{\e}$. Moreover, if
$\wo\geq \Gamma(\e=[30 \alf\sqrt{ \wco} T]^{2/3},T)$ then in the
whole integration range
$\wo\gg\Gamma(\e,T)\Leftrightarrow\wo\tau(\e,T)\gg1$ (see
fig.(\ref{regimesQ}), so that
$\wo\tau/[1+(\wo\tau)^2]\approx1/(\wo\tau)$, see
fig.(\ref{regimesQ}). $Q^{-1}_{\rm{rel}}(\wo,T)$ follows:
\begin{figure}
\begin{center}
\includegraphics[width=7cm]{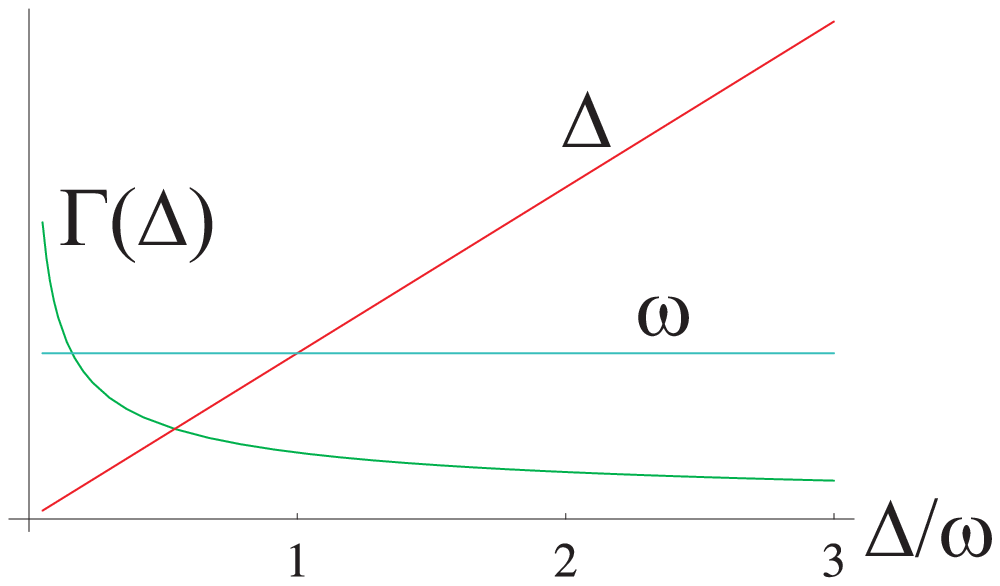}\\
    \caption{(Color online) Evolution with $\Dr$ of the different quantities determining the approximations to be taken in the integrand
 $\wo\tau/[1+(\wo\tau)^2]$ of eq.(\ref{QEsq}).}
    \label{regimesQ}
\end{center}
\end{figure}
\begin{eqnarray}\label{QEsq0}
 Q^{-1}_{\rm{rel}}(\wo,T)&\approx& \frac{P_0\gamma^2}{ET}\int_{[30 \alf\sqrt{ \wco} T]^{2/3}}^{T}d\e\int_{u_{min}}^{1}du\times\nonumber \\
 & \times& \frac{\sqrt{1-u^2}}{u}\frac{\Gamma(\e,T)}{\wo}
\end{eqnarray}

For temperatures $T\gg[30 \alf\sqrt{ \wco} T]^{2/3}$, which holds
for reasonable T and sizes, the integral, which renders a result of
the kind $Q^{-1}_{\rm{rel}}(\wo,T)\approx
(1/\wo)(A\sqrt{T}-BT^{1/3})$, can be approximated by just the first
term, obtaining eq.(\ref{QEsq2}). In any case for completeness we
give the expression for B:
\begin{equation}
 B\approx\frac{500P_0\gamma^{14/3}}{t^2w^{4/3}}\frac{(1+\nu)^{4/3}(1-2\nu)^{4/3}}{E^{11/3}(3-5\nu)}\Bigl(\frac{\rho}{E}\Bigr)^{1/3}
\end{equation}
Also for completeness we give the result for higher temperatures,
although for current sizes the condition $\Gamma(\e=[30 \alf\sqrt{
\wco} T]^{2/3},T)>\wo$ implies values of T above the range of
applicability of the Standard Tunneling Model. Now for some range of
energies $\Gamma(\e,T)>\wo$, and the range of integration is divided
into two regions, one where $\wo\tau\gg1$ and one where the opposite
holds:
\begin{widetext}
\begin{equation}\label{QforbigT}
    Q^{-1}_{\rm{rel}}(\wo,T)=\frac{P_0\gamma^2}{ET}\int_{u_{min}}^{1}du\frac{\sqrt{1-u^2}}{u}\Bigl\{\int_{[30 \alf\sqrt{ \wco} T]^{2/3}}^
{[16\alf\sqrt{\wco}2T/\sqrt{\wo}]^2}dE\,\,
\wo\tau(\e,T)+\int_{[16\alf\sqrt{\wco}2T/\sqrt{\wo}]^2}^{T}dE
\frac{1}{\wo\tau(\e,T)}\Bigr\}
\end{equation}
\end{widetext}
The final result is $Q^{-1}_{\rm{rel}}(\wo,T)\approx
-7P_0\gamma^2\wo/T+A\sqrt{T}/\wo-CT/\wo^2$, with $A$ defined by
eq.(\ref{QEsq2}) and C by:
\begin{equation}
 C\approx\frac{1500P_0\gamma^{6}}{t^3w^{2}}\frac{(1+\nu)^{2}(1-2\nu)^{2}}{E^{3}(3-5\nu)^2}\Bigl(\frac{\rho}{E}\Bigr)^{1/2}
\end{equation}
All the results for $Q^{-1}_{\rm{rel}}(\wo,T)$ have to be multiplied
by the fraction of volume of the resonator presenting amorphous
features, $V_{amorph}/(twL)$.

\section{Derivation of
eqs.(\ref{Hinttors})-(\ref{E0tors})}\label{Tors} To derive the
interaction Hamiltonian, eq.(\ref{Hinttors}), note that in terms of
the deformation gradient matrix $\partial_i u_j$ it must be
$H_{\rm{int}}=\gamma \sigma_z \partial_i u_j $, where the
deformations are caused in this case by twisting of the resonator
about its main axis. The twisting modes correspond $\,\,$\cite{LL59}
to the rotation angle around the longest main axis $\phi$ obeying
the wave equation
\begin{equation}
    C\frac{\partial^2 \phi}{\partial z^2}=\rho I \frac{\partial^2 \phi}{\partial t^2}
\end{equation}
so the variable $\phi$ can be expressed in terms of boson operators
\begin{eqnarray}\label{eqphi}
    \nonumber\phi(0)&=&\sum_k \sqrt{\frac{\hbar}{2\rho twL
    \om_k}}(a_{k}^{\dag}+a_k) \,,\\
    \partial \phi\mid_{z=0}&=& \sqrt{\frac{\hbar}{2\rho twL}}\sum_k \frac{k}{\sqrt{\om_k}}(a_{k}^{\dag}+a_k)
\end{eqnarray}
To obtain in terms of the modes' operators $a_k$ an approximate
expression for $\partial_i u_j $, we relate $\partial u$ to
$\partial \phi$ through the expressions for the free energy of the
rod in terms of both variables,
\begin{widetext}
\begin{equation}
  F_{rod} = \frac{1}{2}\int dz C \left( \frac{\partial \phi}{\partial z}  \right)^2
   = \frac{1}{2}\int dz \int dS 4 \mu \left[ \left( \frac{\partial
   u_x}{\partial z} \right)^2 +
   \left( \frac{\partial u_y}{\partial z} \right)^2 \right] \approx
   \frac{1}{2}\int dz \int dS
8 \mu \left( \frac{\partial u_x}{\partial z} \right)^2
\end{equation}
\end{widetext}
and the approximate relation $\partial u \mid_{z=0}=\partial
u_x/\partial z=\sqrt{C/(8\mu tw)}\partial \phi \mid_{z=0}$ is found.
This relation together with eq.(\ref{eqphi}) lead to the stated
result, eq.(\ref{Hinttors}).

\textbf{Calculation of E$_0$.}To calculate classically the energy
stored of a torsional mode $\phi_j(z,t)=A\sin[(2j-1)\pi
z/(2L)]\sin(\om_j t)$ we just calculate the kinetic energy in a
moment where the elastic energy is zero, for example at time $t=0$.
If an element of mass is originally at position $(x,y,z)$ ($x,y$
transversal coordinates), with a torsion $\phi(z,t)$ it moves to
\begin{widetext}
\begin{equation}\label{vector}
    \vec{r}(t)=\Bigl ( \sqrt{x^2+y^2}\cdot\cos\Bigl[\arccos\frac{x}{\sqrt{x^2+y^2}}+\phi\Bigr],
    \sqrt{x^2+y^2}\cdot\sin\Bigl[\arccos\frac{x}{\sqrt{x^2+y^2}}+\phi\Bigr],z\Bigr )
\end{equation}
\end{widetext}
The kinetic energy at time $t=0$ is
\begin{equation}\label{kin}
    E_{0}=\int_{0}^{L}dz\int_{-t/2}^{t/2}dx\int_{-w/2}^{w/2}dy\cdot\frac{1}{2}\rho
    \Bigl|\frac{d\overrightarrow{r}}{dt}\Bigr|^2_{t=0}
\end{equation}
Substituting the expression for $\vec{r}(t)$ in the integrand, one
arrives at
\begin{widetext}
\begin{equation}\label{Erod}
    E_{0}=\int_{0}^{L}dz\int_{-t/2}^{t/2}dx\int_{-w/2}^{w/2}dy\cdot\frac{1}{2}\rho
    A^2\om_j^2\sin^2\Bigl [\frac{(2j-1)\pi}{2L}z\Bigr
    ](x^2+y^2)=\frac{1}{48}A^2\om_j^2\rho L(t^3w+w^3t)
\end{equation}
\end{widetext}
In terms of the creation and annihilation operators
$\phi_j(z,t)=\frac{\hbar}{2L\rho
I\om_j}(a_{j}^{\dag}+a_j)e^{i(k_jz-\om t)}$, so the mean square of
its amplitude is $\langle \phi_j^2\rangle=A^2/2=\hbar(2n+1)/[2L\rho
I\om_j]$. Substituting this in eq.(\ref{Erod}) the eq.(\ref{E0tors})
for $E_0$ is obtained.

\bibliography{NEMsPRBSi}

\newcommand{\npb}{Nucl. Phys.}\newcommand{\adv}{Adv.
  Phys.}\newcommand{\epl}{Europhys. Lett.}
\begin{thebibliography}{65}
\expandafter\ifx\csname natexlab\endcsname\relax\def\natexlab#1{#1}\fi
\expandafter\ifx\csname bibnamefont\endcsname\relax
  \def\bibnamefont#1{#1}\fi
\expandafter\ifx\csname bibfnamefont\endcsname\relax
  \def\bibfnamefont#1{#1}\fi
\expandafter\ifx\csname citenamefont\endcsname\relax
  \def\citenamefont#1{#1}\fi
\expandafter\ifx\csname url\endcsname\relax
  \def\url#1{\texttt{#1}}\fi
\expandafter\ifx\csname urlprefix\endcsname\relax\def\urlprefix{URL }\fi
\providecommand{\bibinfo}[2]{#2}
\providecommand{\eprint}[2][]{\url{#2}}

\bibitem[{\citenamefont{Seoanez et~al.}(2007)\citenamefont{Seoanez, Guinea, and
  {Castro Neto}}}]{SGN07}
\bibinfo{author}{\bibfnamefont{C.}~\bibnamefont{Seoanez}},
  \bibinfo{author}{\bibfnamefont{F.}~\bibnamefont{Guinea}}, \bibnamefont{and}
  \bibinfo{author}{\bibfnamefont{A.~H.} \bibnamefont{{Castro Neto}}},
  \bibinfo{journal}{Europhys. Lett.} \textbf{\bibinfo{volume}{78}},
  \bibinfo{pages}{60002} (\bibinfo{year}{2007}).

\bibitem[{\citenamefont{Sze}(1981)}]{S81}
\bibinfo{author}{\bibfnamefont{S.}~\bibnamefont{Sze}},
  \emph{\bibinfo{title}{Physics of semiconductor devices}}
  (\bibinfo{publisher}{Wiley-Interscience (New York)}, \bibinfo{year}{1981}).

\bibitem[{\citenamefont{Craighead}(2000)}]{C00}
\bibinfo{author}{\bibfnamefont{H.~G.} \bibnamefont{Craighead}},
  \bibinfo{journal}{Science} \textbf{\bibinfo{volume}{250}},
  \bibinfo{pages}{1532} (\bibinfo{year}{2000}).

\bibitem[{\citenamefont{Cleland}(2002)}]{C02}
\bibinfo{author}{\bibfnamefont{A.~N.} \bibnamefont{Cleland}},
  \emph{\bibinfo{title}{Foundations of Nanomechanics}}
  (\bibinfo{publisher}{Springer (Berlin)}, \bibinfo{year}{2002}).

\bibitem[{\citenamefont{Blencowe}(2004)}]{B04}
\bibinfo{author}{\bibfnamefont{M.}~\bibnamefont{Blencowe}},
  \bibinfo{journal}{Phys. Rep.} \textbf{\bibinfo{volume}{395}},
  \bibinfo{pages}{159} (\bibinfo{year}{2004}).

\bibitem[{\citenamefont{Ekinci and Roukes}(2005)}]{ER05}
\bibinfo{author}{\bibfnamefont{K.~L.} \bibnamefont{Ekinci}} \bibnamefont{and}
  \bibinfo{author}{\bibfnamefont{M.~L.} \bibnamefont{Roukes}},
  \bibinfo{journal}{Rev. Sci. Inst.} \textbf{\bibinfo{volume}{76}},
  \bibinfo{pages}{061101} (\bibinfo{year}{2005}).

\bibitem[{\citenamefont{Scheible and Blick}(2004)}]{SB04}
\bibinfo{author}{\bibfnamefont{D.~V.} \bibnamefont{Scheible}} \bibnamefont{and}
  \bibinfo{author}{\bibfnamefont{R.~H.} \bibnamefont{Blick}},
  \bibinfo{journal}{Appl. Phys. Lett.} \textbf{\bibinfo{volume}{84}},
  \bibinfo{pages}{4632} (\bibinfo{year}{2004}).

\bibitem[{\citenamefont{Rugar et~al.}(2004)\citenamefont{Rugar, Budakian,
  Mamin, and Chui}}]{RBMC04}
\bibinfo{author}{\bibfnamefont{D.}~\bibnamefont{Rugar}},
  \bibinfo{author}{\bibfnamefont{R.}~\bibnamefont{Budakian}},
  \bibinfo{author}{\bibfnamefont{H.}~\bibnamefont{Mamin}}, \bibnamefont{and}
  \bibinfo{author}{\bibfnamefont{B.}~\bibnamefont{Chui}},
  \bibinfo{journal}{Nature} \textbf{\bibinfo{volume}{430}},
  \bibinfo{pages}{329} (\bibinfo{year}{2004}).

\bibitem[{\citenamefont{Mamin and Rugar}(2001)}]{MR01}
\bibinfo{author}{\bibfnamefont{H.~J.} \bibnamefont{Mamin}} \bibnamefont{and}
  \bibinfo{author}{\bibfnamefont{D.}~\bibnamefont{Rugar}},
  \bibinfo{journal}{Appl. Phys. Lett.} \textbf{\bibinfo{volume}{79}},
  \bibinfo{pages}{3358} (\bibinfo{year}{2001}).

\bibitem[{\citenamefont{K.~L.~Ekinci and Roukes}(2004)}]{EHR04}
\bibinfo{author}{\bibfnamefont{X.~M. H.~H.} \bibnamefont{K.~L.~Ekinci}}
  \bibnamefont{and} \bibinfo{author}{\bibfnamefont{M.~L.}
  \bibnamefont{Roukes}}, \bibinfo{journal}{Appl. Phys. Lett.}
  \textbf{\bibinfo{volume}{84}}, \bibinfo{pages}{4469} (\bibinfo{year}{2004}).

\bibitem[{\citenamefont{Hopcroft et~al.}(2007)\citenamefont{Hopcroft, Kim,
  Chandorkar, Melamud, Agarwal, Jha, Bahl, Salvia, Mehta, Lee
  et~al.}}]{hopcroft:013505}
\bibinfo{author}{\bibfnamefont{M.~A.} \bibnamefont{Hopcroft}},
  \bibinfo{author}{\bibfnamefont{B.}~\bibnamefont{Kim}},
  \bibinfo{author}{\bibfnamefont{S.}~\bibnamefont{Chandorkar}},
  \bibinfo{author}{\bibfnamefont{R.}~\bibnamefont{Melamud}},
  \bibinfo{author}{\bibfnamefont{M.}~\bibnamefont{Agarwal}},
  \bibinfo{author}{\bibfnamefont{C.~M.} \bibnamefont{Jha}},
  \bibinfo{author}{\bibfnamefont{G.}~\bibnamefont{Bahl}},
  \bibinfo{author}{\bibfnamefont{J.}~\bibnamefont{Salvia}},
  \bibinfo{author}{\bibfnamefont{H.}~\bibnamefont{Mehta}},
  \bibinfo{author}{\bibfnamefont{H.~K.} \bibnamefont{Lee}},
  \bibnamefont{et~al.}, \bibinfo{journal}{Appl. Phys. Lett.}
  \textbf{\bibinfo{volume}{91}}, \bibinfo{eid}{013505} (\bibinfo{year}{2007}).

\bibitem[{\citenamefont{Dorignac et~al.}(2006)\citenamefont{Dorignac,
  Kalinowski, Erramilli, and Mohanty}}]{DKEM06}
\bibinfo{author}{\bibfnamefont{J.}~\bibnamefont{Dorignac}},
  \bibinfo{author}{\bibfnamefont{A.}~\bibnamefont{Kalinowski}},
  \bibinfo{author}{\bibfnamefont{S.}~\bibnamefont{Erramilli}},
  \bibnamefont{and} \bibinfo{author}{\bibfnamefont{P.}~\bibnamefont{Mohanty}},
  \bibinfo{journal}{Phys. Rev. Lett.} \textbf{\bibinfo{volume}{96}},
  \bibinfo{pages}{186105} (\bibinfo{year}{2006}).

\bibitem[{\citenamefont{LaHaye et~al.}(2004)\citenamefont{LaHaye, Buu,
  Camarota, and Schwab}}]{LBCS04}
\bibinfo{author}{\bibfnamefont{M.~D.} \bibnamefont{LaHaye}},
  \bibinfo{author}{\bibfnamefont{O.}~\bibnamefont{Buu}},
  \bibinfo{author}{\bibfnamefont{B.}~\bibnamefont{Camarota}}, \bibnamefont{and}
  \bibinfo{author}{\bibfnamefont{K.~C.} \bibnamefont{Schwab}},
  \bibinfo{journal}{Science} \textbf{\bibinfo{volume}{304}},
  \bibinfo{pages}{74} (\bibinfo{year}{2004}).

\bibitem[{\citenamefont{Zolfagharkhani
  et~al.}(2005{\natexlab{a}})\citenamefont{Zolfagharkhani, Gaidarzhy, Badzey,
  and Mohanty}}]{ZGBM05}
\bibinfo{author}{\bibfnamefont{G.}~\bibnamefont{Zolfagharkhani}},
  \bibinfo{author}{\bibfnamefont{A.}~\bibnamefont{Gaidarzhy}},
  \bibinfo{author}{\bibfnamefont{R.~L.} \bibnamefont{Badzey}},
  \bibnamefont{and} \bibinfo{author}{\bibfnamefont{P.}~\bibnamefont{Mohanty}},
  \bibinfo{journal}{Phys. Rev. Lett.} \textbf{\bibinfo{volume}{94}},
  \bibinfo{pages}{030402} (\bibinfo{year}{2005}{\natexlab{a}}).

\bibitem[{\citenamefont{Naik et~al.}(2006)\citenamefont{Naik, Buu, LaHaye,
  Armour, Clerk, Blencowe, and Schwab}}]{Netal06}
\bibinfo{author}{\bibfnamefont{A.}~\bibnamefont{Naik}},
  \bibinfo{author}{\bibfnamefont{O.}~\bibnamefont{Buu}},
  \bibinfo{author}{\bibfnamefont{M.~D.} \bibnamefont{LaHaye}},
  \bibinfo{author}{\bibfnamefont{A.~D.} \bibnamefont{Armour}},
  \bibinfo{author}{\bibfnamefont{A.~A.} \bibnamefont{Clerk}},
  \bibinfo{author}{\bibfnamefont{M.~P.} \bibnamefont{Blencowe}},
  \bibnamefont{and} \bibinfo{author}{\bibfnamefont{K.~C.}
  \bibnamefont{Schwab}}, \bibinfo{journal}{Nature}
  \textbf{\bibinfo{volume}{443}}, \bibinfo{pages}{193} (\bibinfo{year}{2006}).

\bibitem[{\citenamefont{Santamore et~al.}(2004)\citenamefont{Santamore,
  Doherty, and Cross}}]{santamore:144301}
\bibinfo{author}{\bibfnamefont{D.~H.} \bibnamefont{Santamore}},
  \bibinfo{author}{\bibfnamefont{A.~C.} \bibnamefont{Doherty}},
  \bibnamefont{and} \bibinfo{author}{\bibfnamefont{M.~C.} \bibnamefont{Cross}},
  \bibinfo{journal}{Phys. Rev. B} \textbf{\bibinfo{volume}{70}},
  \bibinfo{pages}{144301} (\bibinfo{year}{2004}).

\bibitem[{\citenamefont{Martin and Zurek}(2007)}]{martin:120401}
\bibinfo{author}{\bibfnamefont{I.}~\bibnamefont{Martin}} \bibnamefont{and}
  \bibinfo{author}{\bibfnamefont{W.~H.} \bibnamefont{Zurek}},
  \bibinfo{journal}{Phys. Rev. Lett.} \textbf{\bibinfo{volume}{98}},
  \bibinfo{pages}{120401} (\bibinfo{year}{2007}).

\bibitem[{\citenamefont{Jacobs et~al.}(2007)\citenamefont{Jacobs, Lougovski,
  and Blencowe}}]{jacobs:147201}
\bibinfo{author}{\bibfnamefont{K.}~\bibnamefont{Jacobs}},
  \bibinfo{author}{\bibfnamefont{P.}~\bibnamefont{Lougovski}},
  \bibnamefont{and} \bibinfo{author}{\bibfnamefont{M.}~\bibnamefont{Blencowe}},
  \bibinfo{journal}{Phys. Rev. Lett.} \textbf{\bibinfo{volume}{98}},
  \bibinfo{pages}{147201} (\bibinfo{year}{2007}).

\bibitem[{\citenamefont{Wei et~al.}(2006)\citenamefont{Wei, xi~Liu, Sun, and
  Nori}}]{WLSN06}
\bibinfo{author}{\bibfnamefont{L.~F.} \bibnamefont{Wei}},
  \bibinfo{author}{\bibfnamefont{Y.}~\bibnamefont{xi~Liu}},
  \bibinfo{author}{\bibfnamefont{C.~P.} \bibnamefont{Sun}}, \bibnamefont{and}
  \bibinfo{author}{\bibfnamefont{F.}~\bibnamefont{Nori}},
  \bibinfo{journal}{Phys. Rev. Lett.} \textbf{\bibinfo{volume}{97}},
  \bibinfo{pages}{237201} (\bibinfo{year}{2006}).

\bibitem[{\citenamefont{Thompson et~al.}(2007)\citenamefont{Thompson, Zwickl,
  Jayich, Marquardt, Girvin, and Harris}}]{Tetal07}
\bibinfo{author}{\bibfnamefont{J.~D.} \bibnamefont{Thompson}},
  \bibinfo{author}{\bibfnamefont{B.~M.} \bibnamefont{Zwickl}},
  \bibinfo{author}{\bibfnamefont{A.~M.} \bibnamefont{Jayich}},
  \bibinfo{author}{\bibfnamefont{F.}~\bibnamefont{Marquardt}},
  \bibinfo{author}{\bibfnamefont{S.~M.} \bibnamefont{Girvin}},
  \bibnamefont{and} \bibinfo{author}{\bibfnamefont{J.~G.~E.}
  \bibnamefont{Harris}} (\bibinfo{year}{2007}), \eprint{arXiv:0707.1724}.

\bibitem[{\citenamefont{Katz et~al.}(2007)\citenamefont{Katz, Retzker, Straub,
  and Lifshitz}}]{katz:040404}
\bibinfo{author}{\bibfnamefont{I.}~\bibnamefont{Katz}},
  \bibinfo{author}{\bibfnamefont{A.}~\bibnamefont{Retzker}},
  \bibinfo{author}{\bibfnamefont{R.}~\bibnamefont{Straub}}, \bibnamefont{and}
  \bibinfo{author}{\bibfnamefont{R.}~\bibnamefont{Lifshitz}},
  \bibinfo{journal}{Phys. Rev. Lett.} \textbf{\bibinfo{volume}{99}},
  \bibinfo{pages}{040404} (\bibinfo{year}{2007}).

\bibitem[{\citenamefont{Cleland and Roukes}(1999)}]{CR99}
\bibinfo{author}{\bibfnamefont{A.~N.} \bibnamefont{Cleland}} \bibnamefont{and}
  \bibinfo{author}{\bibfnamefont{M.~L.} \bibnamefont{Roukes}},
  \bibinfo{journal}{Sens. Actuators A} \textbf{\bibinfo{volume}{72}},
  \bibinfo{pages}{256} (\bibinfo{year}{1999}).

\bibitem[{\citenamefont{Yasumura et~al.}(2000)\citenamefont{Yasumura, Stowe,
  Chow, Pfafman, Kenny, Stipe, and Rugar}}]{Yetal00}
\bibinfo{author}{\bibfnamefont{K.~Y.} \bibnamefont{Yasumura}},
  \bibinfo{author}{\bibfnamefont{T.~D.} \bibnamefont{Stowe}},
  \bibinfo{author}{\bibfnamefont{E.~M.} \bibnamefont{Chow}},
  \bibinfo{author}{\bibfnamefont{T.}~\bibnamefont{Pfafman}},
  \bibinfo{author}{\bibfnamefont{T.~W.} \bibnamefont{Kenny}},
  \bibinfo{author}{\bibfnamefont{B.~C.} \bibnamefont{Stipe}}, \bibnamefont{and}
  \bibinfo{author}{\bibfnamefont{D.}~\bibnamefont{Rugar}}, \bibinfo{journal}{J.
  Microelectromech. Syst.} \textbf{\bibinfo{volume}{9}}, \bibinfo{pages}{117}
  (\bibinfo{year}{2000}).

\bibitem[{\citenamefont{Evoy et~al.}(2000)\citenamefont{Evoy, Olkhovets,
  Sekaric, Parpia, Craighead, and Carr}}]{Eetal00}
\bibinfo{author}{\bibfnamefont{S.}~\bibnamefont{Evoy}},
  \bibinfo{author}{\bibfnamefont{A.}~\bibnamefont{Olkhovets}},
  \bibinfo{author}{\bibfnamefont{L.}~\bibnamefont{Sekaric}},
  \bibinfo{author}{\bibfnamefont{J.~M.} \bibnamefont{Parpia}},
  \bibinfo{author}{\bibfnamefont{H.~G.} \bibnamefont{Craighead}},
  \bibnamefont{and} \bibinfo{author}{\bibfnamefont{D.~W.} \bibnamefont{Carr}},
  \bibinfo{journal}{Appl. Phys. Lett.} \textbf{\bibinfo{volume}{77}},
  \bibinfo{pages}{2397} (\bibinfo{year}{2000}).

\bibitem[{\citenamefont{Cleland and Roukes}(2002)}]{CR02}
\bibinfo{author}{\bibfnamefont{A.~N.} \bibnamefont{Cleland}} \bibnamefont{and}
  \bibinfo{author}{\bibfnamefont{M.~L.} \bibnamefont{Roukes}},
  \bibinfo{journal}{J. Appl. Phys.} \textbf{\bibinfo{volume}{92}},
  \bibinfo{pages}{2758} (\bibinfo{year}{2002}).

\bibitem[{\citenamefont{Yang et~al.}(2002)\citenamefont{Yang, Ono, and
  Esashi}}]{YOE02}
\bibinfo{author}{\bibfnamefont{J.~L.} \bibnamefont{Yang}},
  \bibinfo{author}{\bibfnamefont{T.}~\bibnamefont{Ono}}, \bibnamefont{and}
  \bibinfo{author}{\bibfnamefont{M.}~\bibnamefont{Esashi}},
  \bibinfo{journal}{J. Microelectromech. Syst.} \textbf{\bibinfo{volume}{11}},
  \bibinfo{pages}{775} (\bibinfo{year}{2002}).

\bibitem[{\citenamefont{Mohanty et~al.}(2002)\citenamefont{Mohanty, Harrington,
  Ekinci, Yang, Murphy, and Roukes}}]{Metal02}
\bibinfo{author}{\bibfnamefont{P.}~\bibnamefont{Mohanty}},
  \bibinfo{author}{\bibfnamefont{D.~A.} \bibnamefont{Harrington}},
  \bibinfo{author}{\bibfnamefont{K.~L.} \bibnamefont{Ekinci}},
  \bibinfo{author}{\bibfnamefont{Y.~T.} \bibnamefont{Yang}},
  \bibinfo{author}{\bibfnamefont{M.~J.} \bibnamefont{Murphy}},
  \bibnamefont{and} \bibinfo{author}{\bibfnamefont{M.~L.}
  \bibnamefont{Roukes}}, \bibinfo{journal}{Phys. Rev. B}
  \textbf{\bibinfo{volume}{66}}, \bibinfo{pages}{085416}
  (\bibinfo{year}{2002}).

\bibitem[{\citenamefont{Ahn and Mohanty}(2003)}]{AM03}
\bibinfo{author}{\bibfnamefont{K.-H.} \bibnamefont{Ahn}} \bibnamefont{and}
  \bibinfo{author}{\bibfnamefont{P.}~\bibnamefont{Mohanty}},
  \bibinfo{journal}{Phys. Rev. Lett.} \textbf{\bibinfo{volume}{90}},
  \bibinfo{pages}{085504} (\bibinfo{year}{2003}).

\bibitem[{\citenamefont{Husain et~al.}(2003)\citenamefont{Husain, Hone, Postma,
  Huang, Drake, Barbic, Scherer, and Roukes}}]{Hetal03}
\bibinfo{author}{\bibfnamefont{A.}~\bibnamefont{Husain}},
  \bibinfo{author}{\bibfnamefont{J.}~\bibnamefont{Hone}},
  \bibinfo{author}{\bibfnamefont{H.~W.~C.} \bibnamefont{Postma}},
  \bibinfo{author}{\bibfnamefont{X.~M.~H.} \bibnamefont{Huang}},
  \bibinfo{author}{\bibfnamefont{T.}~\bibnamefont{Drake}},
  \bibinfo{author}{\bibfnamefont{M.}~\bibnamefont{Barbic}},
  \bibinfo{author}{\bibfnamefont{A.}~\bibnamefont{Scherer}}, \bibnamefont{and}
  \bibinfo{author}{\bibfnamefont{M.~L.} \bibnamefont{Roukes}},
  \bibinfo{journal}{Appl. Phys. Lett.} \textbf{\bibinfo{volume}{83}},
  \bibinfo{pages}{1240} (\bibinfo{year}{2003}).

\bibitem[{\citenamefont{Zolfagharkhani
  et~al.}(2005{\natexlab{b}})\citenamefont{Zolfagharkhani, Gaidarzhy, Shim,
  Badzey, and Mohanty}}]{ZGSBM05}
\bibinfo{author}{\bibfnamefont{G.}~\bibnamefont{Zolfagharkhani}},
  \bibinfo{author}{\bibfnamefont{A.}~\bibnamefont{Gaidarzhy}},
  \bibinfo{author}{\bibfnamefont{S.-B.} \bibnamefont{Shim}},
  \bibinfo{author}{\bibfnamefont{R.}~\bibnamefont{Badzey}}, \bibnamefont{and}
  \bibinfo{author}{\bibfnamefont{P.}~\bibnamefont{Mohanty}},
  \bibinfo{journal}{Phys. Rev. B} \textbf{\bibinfo{volume}{72}},
  \bibinfo{pages}{224101} (\bibinfo{year}{2005}{\natexlab{b}}).

\bibitem[{\citenamefont{Feng et~al.}(2006)\citenamefont{Feng, Zorman,
  Mehregany, and Roukes}}]{FZMR06}
\bibinfo{author}{\bibfnamefont{X.~L.} \bibnamefont{Feng}},
  \bibinfo{author}{\bibfnamefont{C.~A.} \bibnamefont{Zorman}},
  \bibinfo{author}{\bibfnamefont{M.}~\bibnamefont{Mehregany}},
  \bibnamefont{and} \bibinfo{author}{\bibfnamefont{M.~L.} \bibnamefont{Roukes}}
  (\bibinfo{year}{2006}), \eprint{cond-mat/0606711}.

\bibitem[{\citenamefont{Shim et~al.}(2007)\citenamefont{Shim, Chun, Kang, Cho,
  Cho, Park, Mohanty, Kim, and Kim}}]{Mohetal07}
\bibinfo{author}{\bibfnamefont{S.-B.} \bibnamefont{Shim}},
  \bibinfo{author}{\bibfnamefont{J.~S.} \bibnamefont{Chun}},
  \bibinfo{author}{\bibfnamefont{S.~W.} \bibnamefont{Kang}},
  \bibinfo{author}{\bibfnamefont{S.~W.} \bibnamefont{Cho}},
  \bibinfo{author}{\bibfnamefont{S.~W.} \bibnamefont{Cho}},
  \bibinfo{author}{\bibfnamefont{Y.~D.} \bibnamefont{Park}},
  \bibinfo{author}{\bibfnamefont{P.}~\bibnamefont{Mohanty}},
  \bibinfo{author}{\bibfnamefont{N.}~\bibnamefont{Kim}}, \bibnamefont{and}
  \bibinfo{author}{\bibfnamefont{J.}~\bibnamefont{Kim}},
  \bibinfo{journal}{Appl. Phys. Lett.} \textbf{\bibinfo{volume}{91}},
  \bibinfo{pages}{133505} (\bibinfo{year}{2007}).

\bibitem[{\citenamefont{Jimbo and Itao}(1968)}]{JI68}
\bibinfo{author}{\bibfnamefont{Y.}~\bibnamefont{Jimbo}} \bibnamefont{and}
  \bibinfo{author}{\bibfnamefont{K.}~\bibnamefont{Itao}}, \bibinfo{journal}{J.
  Horological Inst. Jpn.} \textbf{\bibinfo{volume}{47}}, \bibinfo{pages}{1}
  (\bibinfo{year}{1968}).

\bibitem[{\citenamefont{Photiadis and Judge}(2004)}]{PJ04}
\bibinfo{author}{\bibfnamefont{D.~M.} \bibnamefont{Photiadis}}
  \bibnamefont{and} \bibinfo{author}{\bibfnamefont{J.~A.} \bibnamefont{Judge}},
  \bibinfo{journal}{Applied Physics Letters} \textbf{\bibinfo{volume}{85}},
  \bibinfo{pages}{482} (\bibinfo{year}{2004}).

\bibitem[{\citenamefont{Zener}(1938)}]{Z38}
\bibinfo{author}{\bibfnamefont{C.}~\bibnamefont{Zener}},
  \bibinfo{journal}{Phys. Rev.} \textbf{\bibinfo{volume}{53}},
  \bibinfo{pages}{90} (\bibinfo{year}{1938}).

\bibitem[{\citenamefont{Lifshitz and Roukes}(2000)}]{LR00}
\bibinfo{author}{\bibfnamefont{R.}~\bibnamefont{Lifshitz}} \bibnamefont{and}
  \bibinfo{author}{\bibfnamefont{M.}~\bibnamefont{Roukes}},
  \bibinfo{journal}{Phys. Rev. B} \textbf{\bibinfo{volume}{61}},
  \bibinfo{pages}{5600} (\bibinfo{year}{2000}).

\bibitem[{\citenamefont{De and Aluru}(2006)}]{de:144305}
\bibinfo{author}{\bibfnamefont{S.~K.} \bibnamefont{De}} \bibnamefont{and}
  \bibinfo{author}{\bibfnamefont{N.~R.} \bibnamefont{Aluru}},
  \bibinfo{journal}{Phys. Rev. B} \textbf{\bibinfo{volume}{74}},
  \bibinfo{pages}{144305} (\bibinfo{year}{2006}).

\bibitem[{\citenamefont{Yang et~al.}(2000)\citenamefont{Yang, Ono, and
  Esashi}}]{YOE00}
\bibinfo{author}{\bibfnamefont{J.}~\bibnamefont{Yang}},
  \bibinfo{author}{\bibfnamefont{T.}~\bibnamefont{Ono}}, \bibnamefont{and}
  \bibinfo{author}{\bibfnamefont{M.}~\bibnamefont{Esashi}},
  \bibinfo{journal}{Appl. Phys. Lett.} \textbf{\bibinfo{volume}{77}},
  \bibinfo{pages}{3860} (\bibinfo{year}{2000}).

\bibitem[{\citenamefont{Liu et~al.}(2005)\citenamefont{Liu, Vignola, Simpson,
  Lemon, Houston, and Photiadis}}]{LVSLHP05}
\bibinfo{author}{\bibfnamefont{X.}~\bibnamefont{Liu}},
  \bibinfo{author}{\bibfnamefont{J.~F.} \bibnamefont{Vignola}},
  \bibinfo{author}{\bibfnamefont{H.~J.} \bibnamefont{Simpson}},
  \bibinfo{author}{\bibfnamefont{B.~R.} \bibnamefont{Lemon}},
  \bibinfo{author}{\bibfnamefont{B.~H.} \bibnamefont{Houston}},
  \bibnamefont{and} \bibinfo{author}{\bibfnamefont{D.~M.}
  \bibnamefont{Photiadis}}, \bibinfo{journal}{J. Appl. Phys.}
  \textbf{\bibinfo{volume}{97}}, \bibinfo{pages}{023524}
  (\bibinfo{year}{2005}).

\bibitem[{\citenamefont{Chu et~al.}(2007)\citenamefont{Chu, Rudd, and
  Blencowe}}]{C07}
\bibinfo{author}{\bibfnamefont{M.}~\bibnamefont{Chu}},
  \bibinfo{author}{\bibfnamefont{R.~E.} \bibnamefont{Rudd}}, \bibnamefont{and}
  \bibinfo{author}{\bibfnamefont{M.~P.} \bibnamefont{Blencowe}}
  (\bibinfo{year}{2007}), \eprint{cond-mat/0705.0015}.

\bibitem[{\citenamefont{Wang et~al.}(2004)\citenamefont{Wang, Ono, and
  Esashi}}]{WOE04}
\bibinfo{author}{\bibfnamefont{D.}~\bibnamefont{Wang}},
  \bibinfo{author}{\bibfnamefont{T.}~\bibnamefont{Ono}}, \bibnamefont{and}
  \bibinfo{author}{\bibfnamefont{M.}~\bibnamefont{Esashi}},
  \bibinfo{journal}{Nanotechnology} \textbf{\bibinfo{volume}{15}},
  \bibinfo{pages}{1851} (\bibinfo{year}{2004}).

\bibitem[{\citenamefont{Liu et~al.}(1999)\citenamefont{Liu, Thompson, { B. E.
  White Jr.}, and Pohl}}]{LTWP99}
\bibinfo{author}{\bibfnamefont{X.}~\bibnamefont{Liu}},
  \bibinfo{author}{\bibfnamefont{E.~J.} \bibnamefont{Thompson}},
  \bibinfo{author}{\bibnamefont{{ B. E. White Jr.}}}, \bibnamefont{and}
  \bibinfo{author}{\bibfnamefont{R.~O.} \bibnamefont{Pohl}},
  \bibinfo{journal}{Phys. Rev. B} \textbf{\bibinfo{volume}{59}},
  \bibinfo{pages}{11767} (\bibinfo{year}{1999}).

\bibitem[{\citenamefont{Anderson et~al.}(1972)\citenamefont{Anderson, Halperin,
  and Varma}}]{AHV72}
\bibinfo{author}{\bibfnamefont{P.}~\bibnamefont{Anderson}},
  \bibinfo{author}{\bibfnamefont{B.}~\bibnamefont{Halperin}}, \bibnamefont{and}
  \bibinfo{author}{\bibfnamefont{C.}~\bibnamefont{Varma}},
  \bibinfo{journal}{Philos. Mag.} \textbf{\bibinfo{volume}{25}},
  \bibinfo{pages}{1} (\bibinfo{year}{1972}).

\bibitem[{\citenamefont{Phillips}(1972)}]{P72}
\bibinfo{author}{\bibfnamefont{W.}~\bibnamefont{Phillips}},
  \bibinfo{journal}{J. Low Temp. Phys.} \textbf{\bibinfo{volume}{7}},
  \bibinfo{pages}{351} (\bibinfo{year}{1972}).

\bibitem[{\citenamefont{Phillips}(1987)}]{P87}
\bibinfo{author}{\bibfnamefont{W.~A.} \bibnamefont{Phillips}},
  \bibinfo{journal}{Rep. Prog. Phys.} \textbf{\bibinfo{volume}{50}},
  \bibinfo{pages}{1657} (\bibinfo{year}{1987}).

\bibitem[{\citenamefont{Esquinazi}(1998)}]{E98}
\bibinfo{editor}{\bibfnamefont{P.}~\bibnamefont{Esquinazi}}, ed.,
  \emph{\bibinfo{title}{Tunneling Systems in Amorphous and Crystalline Solids}}
  (\bibinfo{publisher}{Springer (Berlin)}, \bibinfo{year}{1998}).

\bibitem[{\citenamefont{Yu and Leggett}(1988)}]{YL88}
\bibinfo{author}{\bibfnamefont{C.~C.} \bibnamefont{Yu}} \bibnamefont{and}
  \bibinfo{author}{\bibfnamefont{A.~J.} \bibnamefont{Leggett}},
  \bibinfo{journal}{Comm. Cond. Mat. Phys.} \textbf{\bibinfo{volume}{14}},
  \bibinfo{pages}{231} (\bibinfo{year}{1988}).

\bibitem[{\citenamefont{Esquinazi et~al.}(2004)\citenamefont{Esquinazi, Ramos,
  and K\"onig}}]{ERK04}
\bibinfo{author}{\bibfnamefont{P.}~\bibnamefont{Esquinazi}},
  \bibinfo{author}{\bibfnamefont{M.~A.} \bibnamefont{Ramos}}, \bibnamefont{and}
  \bibinfo{author}{\bibfnamefont{R.}~\bibnamefont{K\"onig}},
  \bibinfo{journal}{J. Low Temp. Phys.} \textbf{\bibinfo{volume}{135}},
  \bibinfo{pages}{27} (\bibinfo{year}{2004}).

\bibitem[{\citenamefont{Shytov et~al.}(2002)\citenamefont{Shytov, Levitov, and
  Beenakker}}]{SLB02}
\bibinfo{author}{\bibfnamefont{A.~V.} \bibnamefont{Shytov}},
  \bibinfo{author}{\bibfnamefont{L.~S.} \bibnamefont{Levitov}},
  \bibnamefont{and} \bibinfo{author}{\bibfnamefont{C.~W.~J.}
  \bibnamefont{Beenakker}}, \bibinfo{journal}{Phys. Rev. Lett.}
  \textbf{\bibinfo{volume}{88}}, \bibinfo{pages}{228303}
  (\bibinfo{year}{2002}).

\bibitem[{\citenamefont{Anderson}(1986)}]{A86}
\bibinfo{author}{\bibfnamefont{A.}~\bibnamefont{Anderson}},
  \bibinfo{journal}{J. Non-Crys. Solids} \textbf{\bibinfo{volume}{85}},
  \bibinfo{pages}{211} (\bibinfo{year}{1986}).

\bibitem[{\citenamefont{Landau and Lifshitz}(1959)}]{LL59}
\bibinfo{author}{\bibfnamefont{L.~D.} \bibnamefont{Landau}} \bibnamefont{and}
  \bibinfo{author}{\bibfnamefont{E.~M.} \bibnamefont{Lifshitz}},
  \emph{\bibinfo{title}{Theory of Elasticity}} (\bibinfo{publisher}{Pergamon
  Press (London)}, \bibinfo{year}{1959}).

\bibitem[{\citenamefont{Arnold et~al.}(1974)\citenamefont{Arnold, Hunklinger,
  Stein, and Dransfeld}}]{Aetal74}
\bibinfo{author}{\bibfnamefont{W.}~\bibnamefont{Arnold}},
  \bibinfo{author}{\bibfnamefont{S.}~\bibnamefont{Hunklinger}},
  \bibinfo{author}{\bibfnamefont{S.}~\bibnamefont{Stein}}, \bibnamefont{and}
  \bibinfo{author}{\bibfnamefont{K.}~\bibnamefont{Dransfeld}},
  \bibinfo{journal}{J. Non Cryst. Sol.} \textbf{\bibinfo{volume}{14}},
  \bibinfo{pages}{192} (\bibinfo{year}{1974}).

\bibitem[{\citenamefont{J\"{a}ckle}(1972)}]{J72}
\bibinfo{author}{\bibfnamefont{J.}~\bibnamefont{J\"{a}ckle}},
  \bibinfo{journal}{Z. Physik} \textbf{\bibinfo{volume}{257}},
  \bibinfo{pages}{212} (\bibinfo{year}{1972}).

\bibitem[{\citenamefont{Mukhopadhyay et~al.}(2005)\citenamefont{Mukhopadhyay,
  Sumbayev, Lorentzen, Kjems, Andreasen, and Besenbacher}}]{Metal05}
\bibinfo{author}{\bibfnamefont{R.}~\bibnamefont{Mukhopadhyay}},
  \bibinfo{author}{\bibfnamefont{V.~V.} \bibnamefont{Sumbayev}},
  \bibinfo{author}{\bibfnamefont{M.}~\bibnamefont{Lorentzen}},
  \bibinfo{author}{\bibfnamefont{J.}~\bibnamefont{Kjems}},
  \bibinfo{author}{\bibfnamefont{P.~A.} \bibnamefont{Andreasen}},
  \bibnamefont{and}
  \bibinfo{author}{\bibfnamefont{F.}~\bibnamefont{Besenbacher}},
  \bibinfo{journal}{Nano Lett.} \textbf{\bibinfo{volume}{5}},
  \bibinfo{pages}{2385} (\bibinfo{year}{2005}).

\bibitem[{\citenamefont{Sharos et~al.}(2004)\citenamefont{Sharos, Raman,
  Crittenden, and Reifenberger}}]{SRCR04}
\bibinfo{author}{\bibfnamefont{L.}~\bibnamefont{Sharos}},
  \bibinfo{author}{\bibfnamefont{A.}~\bibnamefont{Raman}},
  \bibinfo{author}{\bibfnamefont{S.}~\bibnamefont{Crittenden}},
  \bibnamefont{and}
  \bibinfo{author}{\bibfnamefont{R.}~\bibnamefont{Reifenberger}},
  \bibinfo{journal}{Appl. Phys. Lett.} \textbf{\bibinfo{volume}{84}},
  \bibinfo{pages}{4638} (\bibinfo{year}{2004}).

\bibitem[{\citenamefont{Phillips}(1988)}]{P88}
\bibinfo{author}{\bibfnamefont{W.}~\bibnamefont{Phillips}},
  \bibinfo{journal}{Phys. Rev. Lett.} \textbf{\bibinfo{volume}{61}},
  \bibinfo{pages}{2632} (\bibinfo{year}{1988}).

\bibitem[{\citenamefont{Black and Halperin}(1977)}]{BH77}
\bibinfo{author}{\bibfnamefont{J.}~\bibnamefont{Black}} \bibnamefont{and}
  \bibinfo{author}{\bibfnamefont{B.}~\bibnamefont{Halperin}},
  \bibinfo{journal}{Phys. Rev. B} \textbf{\bibinfo{volume}{16}},
  \bibinfo{pages}{2879} (\bibinfo{year}{1977}).

\bibitem[{\citenamefont{Guinea et~al.}(2004)\citenamefont{Guinea, Jalabert, and
  Sols}}]{GJS04}
\bibinfo{author}{\bibfnamefont{F.}~\bibnamefont{Guinea}},
  \bibinfo{author}{\bibfnamefont{R.~A.} \bibnamefont{Jalabert}},
  \bibnamefont{and} \bibinfo{author}{\bibfnamefont{F.}~\bibnamefont{Sols}},
  \bibinfo{journal}{Phys. Rev. B} \textbf{\bibinfo{volume}{70}},
  \bibinfo{pages}{085310} (\bibinfo{year}{2004}).

\bibitem[{\citenamefont{Guinea}(2005)}]{G05}
\bibinfo{author}{\bibfnamefont{F.}~\bibnamefont{Guinea}},
  \bibinfo{journal}{Phys. Rev. B} \textbf{\bibinfo{volume}{71}},
  \bibinfo{pages}{045424} (\bibinfo{year}{2005}).

\bibitem[{\citenamefont{{White Jr.} and Pohl}(1995)}]{WR95}
\bibinfo{author}{\bibfnamefont{B.~E.} \bibnamefont{{White Jr.}}}
  \bibnamefont{and} \bibinfo{author}{\bibfnamefont{R.~O.} \bibnamefont{Pohl}},
  in \emph{\bibinfo{booktitle}{Mat. Res. Soc. Symp.}} (\bibinfo{year}{1995}),
  vol. \bibinfo{volume}{356}, p. \bibinfo{pages}{567}.

\bibitem[{\citenamefont{Nozieres and Pines}(1999)}]{NP99}
\bibinfo{author}{\bibfnamefont{P.}~\bibnamefont{Nozieres}} \bibnamefont{and}
  \bibinfo{author}{\bibfnamefont{D.}~\bibnamefont{Pines}},
  \emph{\bibinfo{title}{The Theory of Quantum Liquids (Third Edition)}}
  (\bibinfo{publisher}{Perseus Books (Cambridge, Massachusets)},
  \bibinfo{year}{1999}).

\bibitem[{\citenamefont{Leggett et~al.}(1987)\citenamefont{Leggett,
  Chakravarty, Dorsey, Fisher, Garg, and Zwerger}}]{Letal87}
\bibinfo{author}{\bibfnamefont{A.~J.} \bibnamefont{Leggett}},
  \bibinfo{author}{\bibfnamefont{S.}~\bibnamefont{Chakravarty}},
  \bibinfo{author}{\bibfnamefont{A.~T.} \bibnamefont{Dorsey}},
  \bibinfo{author}{\bibfnamefont{M.~P.~A.} \bibnamefont{Fisher}},
  \bibinfo{author}{\bibfnamefont{A.}~\bibnamefont{Garg}}, \bibnamefont{and}
  \bibinfo{author}{\bibfnamefont{W.}~\bibnamefont{Zwerger}},
  \bibinfo{journal}{Rev. Mod. Phys.} \textbf{\bibinfo{volume}{59}},
  \bibinfo{pages}{1} (\bibinfo{year}{1987}).

\bibitem[{\citenamefont{Kehrein and Mielke}(1996)}]{KM96}
\bibinfo{author}{\bibfnamefont{S.}~\bibnamefont{Kehrein}} \bibnamefont{and}
  \bibinfo{author}{\bibfnamefont{A.}~\bibnamefont{Mielke}},
  \bibinfo{journal}{Phys. Lett. A} \textbf{\bibinfo{volume}{219}},
  \bibinfo{pages}{313} (\bibinfo{year}{1996}).

\bibitem[{\citenamefont{Weiss}(1999)}]{W99}
\bibinfo{author}{\bibfnamefont{U.}~\bibnamefont{Weiss}},
  \emph{\bibinfo{title}{Quantum Dissipative Systems}}
  (\bibinfo{publisher}{World Scientific (Singapore)}, \bibinfo{year}{1999}).

\bibitem[{\citenamefont{Guinea}(1985)}]{G85}
\bibinfo{author}{\bibfnamefont{F.}~\bibnamefont{Guinea}},
  \bibinfo{journal}{Phys. Rev. B} \textbf{\bibinfo{volume}{32}},
  \bibinfo{pages}{4486} (\bibinfo{year}{1985}).

\end{thebibliography}
\end{document}